\documentclass[sigconf]{acmart}

\usepackage{multirow}
\usepackage[normalem]{ulem}
\usepackage{tikz}
\usepackage{amsmath}
\usepackage{graphicx}
\usepackage{booktabs}
\usepackage{array}
\usepackage{caption}
\usepackage{xcolor,colortbl}
\usepackage{xspace}
\usepackage{makecell}
\usepackage{comment}
\usepackage{subcaption}
\usepackage{bm}
\usepackage{amsfonts}
\usepackage{xurl}

\acmYear{2026}\copyrightyear{2026}
\setcopyright{cc}
\setcctype[4.0]{by}
\acmConference[ASIA CCS '26]{ACM Asia Conference on Computer and Communications Security}{June 1--5, 2026}{Bangalore, India}
\acmBooktitle{ACM Asia Conference on Computer and Communications Security (ASIA CCS '26), June 1--5, 2026, Bangalore, India}
\acmDOI{10.1145/3779208.3785264}
\acmISBN{979-8-4007-2356-8/26/06}

\newcommand{\eg}{e.g.,\xspace}
\newcommand{\etal}{et al.\xspace}
\newcommand{\ie}{i.e.,\xspace}

\newcommand{\sysname}{{Nimai}\xspace}
\newcommand{\nbase}{{Nimai-S}\xspace}
\newcommand{\nrandom}{{Nimai-C}\xspace}
\newcommand{\nimai}{{Nimai}\xspace}
\newcommand{\bgp}{{BGP}\xspace}
\newcommand{\iot}{{IoT}\xspace}
\newcommand{\tor}{{Tor}\xspace}
\newcommand{\microsoft}{{MS-Malware}\xspace}
\newcommand{\cookie}{{Cookie}\xspace}
\newcommand{\BODMAS}{{BODMAS}\xspace}
\newcommand{\bodmas}{{BODMAS}\xspace}
\newcommand{\nprintml}{{nPrintML}\xspace}
\newcommand{\smote}{{SMOTE}\xspace}
\newcommand{\mcccr}{{MC-CCR}\xspace}
\newcommand{\tvae}{{TVAE}\xspace}
\newcommand{\ctabgan}{{CTAB-GAN+}\xspace}
\newcommand{\tabsyn}{{TabSyn}\xspace}
\newcommand{\tabddpm}{{TabDDPM}\xspace}
\newcommand{\great}{{GReaT}\xspace}
\newcommand{\realtabformer}{{REaLTabFormer}\xspace}

\newcommand{\para}[1]{{\xspace{} \bf \noindent {#1.} \hspace{6pt}}}
\newcommand{\paraq}[1]{{\xspace{} \bf \noindent #1 \hspace{6pt}}}

\newenvironment{packed_itemize}{
\begin{list}{\labelitemi}{\leftmargin=1.em}
  \setlength{\itemsep}{0pt}
  \setlength{\parskip}{0pt}
  \setlength{\parsep}{0pt}
  \setlength{\headsep}{0pt}
  \setlength{\topskip}{0pt}
  \setlength{\topmargin}{0pt}
  \setlength{\topsep}{0pt}
  \setlength{\partopsep}{0pt}
}{\end{list}}

\newcolumntype{"}{!{\vrule width 1pt}}
\def\hlinewd#1{%
\noalign{\ifnum0=`}\fi\hrule \@height #1 \futurelet
\reserved@a\@xhline}

\begin{document}

\title{Taming Data Challenges in ML-based Security Tasks Using Generative AI}

\author{Shravya Kanchi}
\affiliation{
  \institution{Virginia Tech}
  \country{shravya@vt.edu}
}

\author{Neal Mangaokar}
\affiliation{
  \institution{University of Michigan, Ann Arbor}
  \country{nealmgkr@umich.edu}
}

\author{Aravind Cheruvu}
\affiliation{
  \institution{Virginia Tech}
  \country{acheruvu@vt.edu}
}

\author{Sifat Muhammad Abdullah}
\affiliation{
  \institution{Virginia Tech}
  \country{sifat@vt.edu}
}

\author{Shirin Nilizadeh}
\affiliation{
  \institution{University of Texas at Arlington}
  \country{shirin.nilizadeh@uta.edu}
}

\author{Atul Prakash}
\affiliation{
  \institution{University of Michigan, Ann Arbor}
  \country{aprakash@umich.edu}
}

\author{Bimal Viswanath}
\affiliation{
  \institution{Virginia Tech}
  \country{vbimal@vt.edu}
}

\renewcommand{\shortauthors}{Kanchi et al.}

\begin{abstract}
 Machine learning-based supervised classifiers are widely used for security tasks, and their improvement has been largely focused on algorithmic advancements. \textit{Data challenges} that negatively impact the performance of these classifiers have received limited attention. We address the following research question: \textit{Can developments in Generative AI (GenAI) address data challenges and improve classifier performance?} We propose augmenting training datasets with synthetic data generated using GenAI techniques to improve classifier generalization. We evaluate this approach across 7 diverse security tasks using 6 state-of-the-art GenAI methods and introduce a novel GenAI scheme called \sysname{} that enables highly controlled data synthesis. We find that GenAI techniques can significantly improve the performance of security classifiers, achieving improvements of up to 32.6\% even in severely data-constrained settings (only $\sim $180 training samples). Furthermore, we demonstrate that GenAI can facilitate rapid adaptation to concept drift post-deployment, requiring minimal labeling in the adjustment process. Despite successes, our study finds that some GenAI schemes struggle to initialize (train and produce data) on certain security tasks. We also identify characteristics of specific tasks, such as noisy labels, overlapping class distributions, and sparse feature vectors, which hinder performance boost using GenAI. We believe that our study will drive the development of future GenAI tools designed for security tasks.

\end{abstract}

\keywords{generative AI, synthetic data, cybersecurity, security classification}

\ccsdesc[500]{Security and privacy~Systems security}
\ccsdesc[300]{Computing methodologies~Classification and regression trees}

\maketitle

\section{Introduction}
\label{sec:intro}

\noindent The application of machine learning (ML) to counteract cybersecurity threats is becoming increasingly prevalent. A widely adopted approach involves creating ML-based \textit{security classifiers} through supervised learning to identify or categorize threats. These classifiers serve to identify or categorize malicious users or behaviors on online platforms~\cite{xu2021deep}, malicious software~\cite{yang2021bodmas}, network breaches~\cite{mirsky5kitsune}, harmful network traffic/devices/entities~\cite{tekiner2022lightweight, cicids2017}, software vulnerabilities~\cite{zhou2019devign}, and web security hazards~\cite{yang2022wtagraph}. These defense mechanisms emphasize lowering error rates (false alarms), enhancing generalization to unfamiliar test scenarios, and adding robustness against adaptive attackers. In pursuing these goals, primary attention has been paid to addressing the \textit{algorithmic challenges}, which pertain to developing effective ML models and feature engineering techniques. However, the \textit{\textbf{data challenges}} that affect the practical effectiveness of these defenses have received limited attention, resulting in significant limitations in further advancing performance. Our examination uncovers the following crucial data challenges in security (Section~\ref{sec:data-challenges-limited}): significant class imbalance, insufficient representation of attack patterns, inadequate training samples, high-dimensional features, and concept drift~\cite{yang2021cade}.

\begin{figure}[t]
    \centering
    \includegraphics[height=0.4\textwidth]{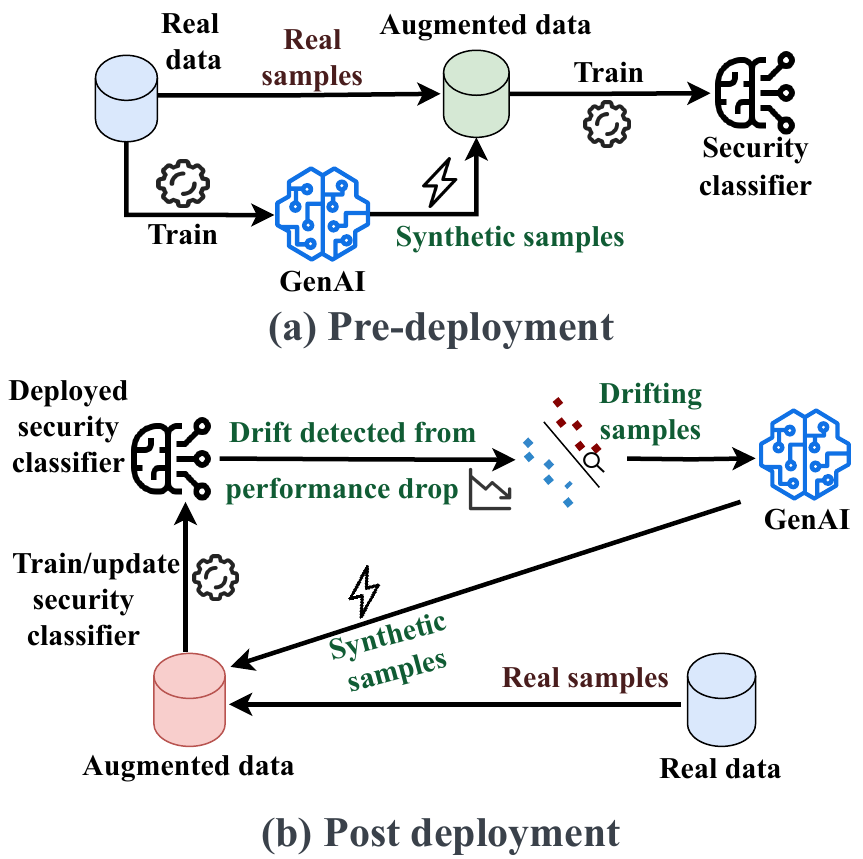}
    \caption{Overview of data augmentation using GenAI.}
    \Description{A diagram showing the process of data augmentation using generative AI. It illustrates how original data is input into a generative model, which produces synthetic data that is then combined with the original dataset to improve classifier performance.}
    \label{fig:nimai-overview}
\end{figure}

In this study, we explore how recent advancements in AI can alleviate data-related challenges to enhance classification performance. This leads to our primary research inquiry: \textit{What are the challenges and opportunities associated with employing Generative Artificial Intelligence (GenAI) to address data challenges in machine learning (ML)-based security applications?} GenAI models are capable of learning the distribution of a dataset and generating varied synthetic instances~\cite{sanh2019distilbert}. A pivotal concept is utilizing carefully crafted synthetic data from GenAI to augment the existing training dataset of a security classifier, thereby enhancing generalization capabilities. Figure~\ref{fig:nimai-overview} (a) depicts the process of leveraging GenAI for data augmentation to refine performance before the classifier's deployment. We specifically concentrate on \textbf{\textit{tabular data tasks}}, as this represents a prevalent methodology for developing ML-based security classifiers. Examples of such tasks are presented in Sections~\ref{sec:measurement-study} and~\ref{sec:tasks-metrics}. Employing GenAI to tackle data challenges in security is difficult for multiple reasons:

\begin{packed_itemize}
\item \textit{Data synthesis in the security domain is inherently harder.} Unlike other domains, \eg computer vision~\cite{lu2022generative, ma2022comprehensive, chlap2021review}, data in the security domain is inherently ``adversarial'' and captures behavior of an entity aiming to evade defenses. Thus, defenders typically have less knowledge of attack data distributions than of benign data. A GenAI scheme has to work with biased and limited attack data, as well as data with noisy samples (erroneous labels).
\item \textit{Existing state-of-the-art GenAI models for tabular data are not tailor-made for security tasks.} Several GenAI models for tabular data have emerged recently, \eg{} 
\tvae~\cite{tvae/ctgan}, \ctabgan~\cite{ctabgan+}, \tabddpm~\cite{tabddpm}, \tabsyn~\cite{tabsyn}, \realtabformer~\cite{solatorio2023realtabformer} and \great \cite{GreaT}. No study has systematically tested these GenAI schemes on security tasks. It is unclear how these GenAI schemes (for tabular data) can adapt to the data challenges in security.
\item \textit{Instantiating a GenAI model for diverse security tasks is challenging.} Widely varying feature dimensionality, multi-class settings, and class imbalances can raise scalability issues or even model collapse during training.
\item \textit{Existing GenAI schemes for tabular data do not offer controlled data generation targeting specific regions of data manifold.} Uncontrolled data synthesis that simply creates many samples broadly mimicking the real data may not always alleviate complex biases within a class.
\item \textit{A GenAI scheme for security should not only address data challenges before deployment, but also after deployment.} Since defenders and attackers are perpetually in an arms race, security classifiers are prone to drifting data or \textit{concept drift} at test time~\cite{andresini2021insomnia}.  
It is unclear how GenAI schemes can help with recovery from concept drift with minimal data labeling effort.
\end{packed_itemize}

We perform the first systematic investigation to reveal the difficulties and advantages of applying GenAI within our specific problem domain. Our contributions are as follows:
\begin{packed_itemize}
\item We start by highlighting key data challenges impacting development of security classifiers. We review 35 papers published in top security venues. We find 32/35 papers report at least one data challenge negatively impacting their classifier performance. We use our findings to select 7 security tasks suffering from these data challenges for evaluation.  This includes diverse tasks: malware classification~\cite{yang2021bodmas, ahmadi2016novel, tekiner2022lightweight}, OS fingerprinting~\cite{holland2021new}, BGP hijacking detection~\cite{testart2019profiling}, Tor-based malware traffic detection~\cite{dodia2022exposing}, and web cookie privacy classification~\cite{bollinger2022automating}.

\item We present the first systematic study of the effectiveness of SoTA GenAI schemes (designed for tabular data) to address data challenges and boost performance of security classifiers. We evaluate 6 SoTA GenAI schemes on our 7 security tasks. This includes GenAI models based on VAEs, LLMs, GANs and Diffusion models. We use synthetic data from GenAI schemes to augment training data for security tasks and boost performance.

\item We identify a missing capability in existing GenAI schemes (for tabular data)---none provide the control to target specific regions of the data manifold to alleviate complex biases. To fill this gap, we propose \textbf{\sysname}, a novel VAE-based scheme, which uses a discrete latent space to enable controlled sample-conditioned generations. Given an existing sample, \sysname can generate samples in its vicinity, providing more fine-grained control to the defender to address in-class bias issues. Our analysis shows that carefully crafted synthetic data improves classifier generalization by mitigating biases in the data.

\item Our analysis identifies both significant opportunities and challenges with using GenAI.\\
\textbf{Opportunities before deployment:} This part focuses on mitigating data challenges before deployment of the classifier. We measure the performance gain (or degradation) from data augmentation compared to using only real training data. We obtain two key findings: (1) Existing GenAI schemes, among \tvae, \ctabgan and \tabddpm demonstrate performance improvement in 2 to 4 tasks, \eg with up to 14.8\% gain in the BGP hijacking detection task. (2) Our new scheme, \sysname which offers highly controlled data generation, achieved the best performance gain in 3/4 tasks.
In an extremely data-challenged BGP hijacking detection task ($<200$ training samples), \sysname outperformed all baselines with up to a 32.6\% gain.  \textit{Our findings demonstrates the significant untapped potential of GenAI in security.}

\textbf{Opportunities after deployment:} We evaluate the potential of GenAI schemes in augmenting data to rapidly \textit{recover} from concept drift scenarios, \ie after deployment. When drift occurs in testing data, significant labeling effort is required to update the classifier based on the drifted set. We design a hybrid approach using \sysname that combines both controlled (sample-conditioning) and uncontrolled (class-conditioning) data synthesis schemes to generate samples. Using this hybrid approach, synthetic samples can be rapidly generated using only a few labeled samples. Figure~\ref{fig:nimai-overview} (b) illustrates the application setting. Using a case study of Portable Executable (PE) malware classification, we show \sysname can quickly recover from concept drift with up to 60.4\% boost in performance by only labeling as few as 64 new samples, once drift is detected. 

\textbf{Challenges:} Our analysis also revealed significant challenges with using GenAI: (1) We discover fundamental challenges with instantiating (training and generating data) existing GenAI models on our diverse tasks. 5/6 existing GenAI schemes failed to instantiate in at least one of the 7 tasks. For example, the LLM-based GenAI scheme called \great~\cite{GreaT} failed to generate data in 6/7 tasks due to scalability issues. (2) We identify 3 tasks with shared challenges of limited or biased data, and added complexities for GenAI: noisy labels, sparse features, and class overlap. Most GenAI schemes, including our new approach, fail to improve performance for these tasks, highlighting directions for future work.

\end{packed_itemize}

We hope that our study will help to develop GenAI tools for security classification that address all identified challenges. Section~\ref{sec:conclusion-future-work} presents specific directions for future work.
Our source code, datasets, and models are made available.\footnote{\url{https://github.com/secml-lab-vt/taming-data-challenges-security-ml.git}}

\section{Data Challenges in Security Defenses}
\label{sec:measurement-study}
\noindent
We conduct a study to characterize data challenges faced by researchers in ML-based cybersecurity tasks. 
We review 35 papers~\footnote{Listed in Table~\ref{data-challenges-tasks} (Appendix).} published in leading venues (\eg IEEE S\&P, USENIX Security and IMC) that developed at least one security classifier. The articles are semi-automatically selected based on their relevance to ML in security research.\footnote{{Our paper selection methodology is detailed in Appendix~\ref{appendix:measurement-paper-selection-method}}.} 
We find that 32/35 articles explicitly reported at least one data challenge that negatively affected classifier performance. We identify the following broad data challenges:

\para{1) Limited or biased training data (20/35 papers)}
\label{sec:data-challenges-limited}
\noindent This includes the following subcategories: 

\textit{a) Imbalanced class distribution (10/35 papers).} Certain classes in the training set have a disproportionately lower number of samples compared to others, which degrades classifier performance~\cite{cidon2019high,schuppen2018fanci,xu2019anatomy,barbero2022transcending, bollinger2022automating}. In security, commonly, the ``attack'' class has limited samples compared to the benign class, \eg for malware detection, due to the challenges in sourcing attack data.
\textit{b) Under-represented in-class attack patterns (9/35 papers).} Training dataset lacks enough instances of the different attack patterns that may arise~\cite{xu2021deep, ho2019detecting, schuppen2018fanci, jordaney2017transcend, han2021sigl}. Under-represented regions can exist within a class.

\textit{c) Insufficient number of samples to fit a high-quality classifier (8/35 papers).} Labeling or feature extraction processes are expensive in terms of computing effort, human effort (requires domain expertise), or other required resources. This is especially true in security where domain expertise is required to analyze software binary or run-time behavior to assign labels. This limits the amount of data collected to train an effective classifier~\cite{banescu2017predicting, jordaney2017transcend, downing2021deepreflect, thirumuruganathan2022siraj, zhou2019devign}. 

\textit{d) High dimensionality of features (3/35).} As feature size increases, the number of samples needed for training increases exponentially, due to the `curse of dimensionality'~\cite{bengio2000taking}. Such tasks tend to underperform, when the training data is limited~\cite{xu2021deep, yang2021cade, park2019empirical}. 

\para{2) Concept drift (14/35 papers)}The classifier generalizes poorly to a test set distribution that has evolved and ``drifted'' away from the training distribution. Attackers are known to adapt over time to bypass defenses, leading to concept drift~\cite{jordaney2017transcend, han2021sigl, yang2021cade, yang2022wtagraph, barbero2022transcending}.

\para{3) Adversarial examples (6/35 papers)}Attackers exploit a core vulnerability of ML models by perturbing samples in the \textit{problem space} (\eg the malware binary to bypass classifiers)~\cite{chen2020training, downing2021deepreflect, chen2021cost, jan2020throwing, bollinger2022automating}. For example, Pierazzi et al.~\cite{pierazzi2020intriguing} identified problem-space transformations for Android malware to create evasive malware samples. Training data lack samples that capture such adversarial behavior, leading to degraded performance against adversarial inputs. 

In this work, we focus mainly on addressing the challenge of limited or biased training data, which is the most common challenge in our measurement study (Section~\ref{sub-sec:genai-eval}). In addition, we explore preliminary directions to rapidly recover from concept drift (Section~\ref{sub-sec:concept-drift-evaluation}). We leave adversarial examples for future work.

\section{Background and Related Work}
\subsection{Problem setting and threat model}

\para{Security tasks and tabular data}We study ML-based security classification tasks that use \textit{tabular data.} Our analysis in Section~\ref{sec:measurement-study} found that 28/35 articles utilize tabular data for ML-based security classification, highlighting its prevalence.
A tabular data sample can include $p$ continuous random variables: ${c_1,...,c_{p}}$ and/or $q$ discrete (multinomial) random variables ${d_1,...,d_q}$ that follow an unknown joint distribution. All training/validation/testing samples are encoded as tabular features. This is usually done by transforming the data from the \textit{problem space} (original input space), such as a malware binary, into the \textit{feature space}, \eg a tabular feature, via static or dynamic analysis of the malware binary. 
We explore GenAI techniques that operate in a tabular feature space for training and generating data. Our focus is on supervised classifiers that need labeled datasets, covering binary and multi-class tasks. Tabular data can represent `attack' and `benign' samples, \eg malware binary vs benign binary, or only attack samples, \eg malware family attribution. Any ML model is applicable for classifier development.

\para{Defender and threat model}The \textit{defender} builds an ML-based security classifier to counter a cybersecurity threat. The task suffers from one or more of the data challenges discussed in Section~\ref{sec:measurement-study}. The defender has access to a dataset to train and validate the performance of the security task. The existing training dataset is called the \textit{real dataset}. Defender aims to use a GenAI tool to improve the performance of their security classifier before deployment and to maintain the performance after deployment, by updating the model over time. The GenAI tool is trained on the real data and then used to augment the real dataset with synthetic data to mitigate data challenges. The defender does not aim to re-design the security classifier, \ie the focus is only on improving performance by addressing the training data challenges. We discuss related work that tightly integrates GenAI schemes into the design of the security classifier in Section~\ref{sec:related-work}, which is not our focus. Prior work has also studied privacy-preserving synthetic data generation for ML tasks~\cite{stadler2022synthetic}, to facilitate open data sharing while balancing privacy and utility.
\textit{However, privacy is not the focus of this study due to the substantial challenges in improving and sustaining classifier performance, with data sharing being a non-goal.} We see it as complementary and a potential avenue for privacy-focused research.

\subsection{Research questions}
\label{sec:goals}
\noindent Our contributions are based on the following key questions:

\paraq{1) What are the challenges with instantiating a GenAI model (designed for tabular data) to generate synthetic data for diverse security tasks?}Extremely skewed and limited real data can present various challenges in training a GenAI tool and generating effective synthetic samples. For example, models can fail to train and converge, result in model collapse during training, or generate poor-quality synthetic samples. Incorrectly estimated hyper-parameters can also derail the GenAI training process. Note, we consider SOTA GenAI tools for tabular data, comprising of diverse models, based on Diffusion, LLMs, GANs and VAEs. Our analysis highlights the effectiveness of diverse model families to learn complex security data distributions.

\paraq{2) Can generating synthetic data using existing SOTA GenAI tools, and training a classifier on the augmented data boost the performance, compared to training only on real data?}We investigate the potential of the existing SOTA GenAI tools (designed for tabular data). We hypothesize that complementing real data with synthetic samples created using advanced GenAI tools can improve classification performance. Note that existing GenAI tools are not tailor-made for security tasks, and no study has systematically tested these SOTA GenAI tools for security tasks. We also compare GenAI tools with traditional data augmentation strategies such as SMOTE.

A reasonable question about our approach is how GenAI can help when training data is biased/limited.
Our insight is that GenAI models that incorporate certain strategies can be trained on biased data to still produce effective synthetic samples. For example, we evaluate \ctabgan~\cite{ctabgan+}, a GAN-based GenAI scheme. \ctabgan uses strategies to prioritize more occurrence of rare samples (\eg underrepresented attack patterns) in training set. It also uses special loss terms to weight rare samples during training. Conditional vectors are also used to guide the generator to generate synthetic samples for a specific class. Similarly, VAE-based GenAI models which use powerful priors over the latent space~\cite{stocksieker2024data, fajardo2018vos, kim2024tvariational}, can produce realistic variations of rare samples in the training set. That said, we do not expect all GenAI schemes to handle biased datasets equally well. Therefore, our work aims to provide a balanced perspective---both opportunities and challenges with using GenAI for security data augmentation.

\paraq{3) Can highly controlled synthetic data generation provide performance improvements for security tasks?}Existing Gen- AI tools provide limited control over the generation process--- methods only allow generation conditioned on a class label. Such an uncontrolled synthesis that broadly mimics the real data is unlikely to capture complex biases within a class. 
We propose a novel GenAI scheme named \textbf{\sysname{}} that uses a VAE to facilitate highly controlled data generation. Conditioning on real samples, the tool provides greater control to correct biases and improve generalization.

\paraq{4) Can synthetic data from GenAI tools help with rapidly recovering from concept drift scenarios after deployment?}We leverage our controlled data synthesis tool, \sysname{} to enable fast recovery from concept drift scenarios. Given a limited number of labeled samples from the drifted distribution, \sysname{} can generate new synthetic samples that capture the drift. 

\subsection{Related work}
\label{sec:related-work}
\para{Synthetic data for security}Compared to other domains, limited work in security has used GenAI tools for data augmentation to \textit{improve performance}. We do not target privacy-preserving data synthesis. Security work using synthetic data can be categorized as follows: \textit{(1) Data augmentation using GenAI methods that are tightly integrated with the classifier.} This includes work by Jan et al.~\cite{jan2020throwing}, where they developed a GAN-based data augmentation scheme for web bot detection. Xu et. al~\cite{xu2020toward} developed a VAE-based data augmentation scheme for intrusion detection. 
VGX~\cite{nong2024vgx} use LLMs to generate textual features retrieved from code samples to boost the performance of software vulnerability detection systems.
These methods are tightly coupled with the security classifier and are harder to generalize to other tasks. \textit{(2) Off-the-shelf use of GenAI models from vision.} Instead of using GenAI models designed specifically for tabular data, researchers have applied GAN models from the vision domain for malicious traffic detection~\cite{hao2021producing, meghdouri2021controllable}. \textit{(3) GenAI methods to generate time-series data.} Gong et. al~\cite{gong2022surakav} designed a GAN to create time series data for website fingerprinting detection. Taylor et. al~\cite{taylor2016anomaly} generate attack data by performing perturbations in the time series sequences for Controller Area Network attacks. STAN\cite{xu2021stan} is a Gaussian Mixture Model proposed to generate multivariate time series network traffic data for cybersecurity tasks. 
\textit{(4) Methods like SMOTE.} Prior work extensively uses simple linear interpolation strategies to create synthetic data, including the SMOTE family and its variants~\cite{smote,mc-ccr,adasyn, krawczyk2019radial}. Methods like SMOTE are known to perform poorly on high-dimensional tasks~\cite{blagus2013smote}, common to security. We compare GenAI with SMOTE-based methods.

\para{Synthetic data in other domains}Computer vision domain has well-established pipelines for data augmentation that are built into popular ML frameworks~\cite{pytorch-vision-transforms}. 
This includes image transformation schemes (\eg cropping, rotation). 
Qui et al.~\cite{qiu2021deepsweep} apply image transformation-based data augmentation to protect DNNs from backdoor attacks in computer vision tasks.
Generative models such as GANs have been extensively used to improve performance of image classifiers, in the fields of agriculture~\cite{lu2022generative}, medicine~\cite{chlap2021review} and robotics~\cite{ma2022comprehensive}.
The networking community has done work to synthetically generate CDN caches~\cite{sabnis2021tragen}, BGP network configurations~\cite{bahnasy2020deepbgp} and network traffic~\cite{lin2020using, yin2022practical, jiang2024netdiffusion}. 

\section{Data-challenged Security Tasks}
\label{sec:tasks-metrics}
\noindent
To address Section~\ref{sec:goals}'s research questions, we select 7 ML-based security classification tasks. This includes 3 binary and 4 multi-class classification tasks. These tasks were selected because they (a) use tabular data for learning, (b) demonstrate degraded classification performance, (c) exhibit one or more of the data challenges identified in Section~\ref{sec:measurement-study}, and (d) have publicly released data and source code, allowing us to reproduce their work.  \textit{All 7 tasks use real-world threat data, and not synthetic data, making our findings applicable to real-world scenarios.}
We follow the original authors' recommended classification schemes. Notably, traditional ML models (e.g., tree-based) often outperform deep learning in these security tasks. Recall that improving performance by using more advanced ML schemes is not our focus. Table \ref{tab:table2_stats} summarizes the tasks.

\para{\bgp ~\cite{testart2019profiling}}A binary classification task, in which Autonomous Systems (AS) are classified as either serial hijackers or non-hijackers. Serial hijackers are adversaries that exploit vulnerabilities in the BGP protocol to repeatedly hijack address prefixes over a long period for malicious purposes. Each data sample contains 52 features covering 8 categories, \eg prefix origination behavior, prefix visibility, longevity of prefix announcements, and address space fragmentation. This is an extremely data-challenged task that suffers from limited or biased training data challenges (a)-(c) (Section~\ref{sec:data-challenges-limited}). The dataset includes just 17 serial hijacker samples and 163 benign samples, used to train an Extremely Randomized Trees classifier~\cite{geurts2006extremely}. The model achieves only a 60.5\% F-score on the attack class. We focus on this class, as the benign class already achieves a 98.53\% F-score with little room for improvement.

\para{\tor~\cite{dodia2022exposing}}A binary classification task that labels network traffic as either Tor-based malware or benign. This is developed using a 175 sized feature vector based on website fingerprinting features from Haze \etal~\cite{hayes2016k}, capturing connection-level traffic characteristics \eg packet ordering statistics. The training dataset has 1,183 malware and 6,932 benign samples used to train an XGBoost classifier which achieves only a 43.9\% F-score on the attack class. However, like the \bgp task, we focus on the attack class since the benign class F-score class is already high at 96\%. This task suffers from limited or biased training data challenges (a)-(c) (Section~\ref{sec:data-challenges-limited}). 

\begin{table}[t]
\centering
\setlength{\tabcolsep}{2pt}
\setlength\extrarowheight{1.5pt}
\begin{tabular}{l"c|ccc|c|c}
\multirow{2}{*}{\textbf{Task}} & \multirow{2}{*}{\textbf{Features}} & \multicolumn{3}{c|}{\textbf{Number of samples}}                                                      & \multirow{2}{*}{\textbf{\begin{tabular}[c]{@{}c@{}}Class \\ size\end{tabular}}} & \multirow{2}{*}{\textbf{F-score}} \\ \cline{3-5}
                               &                                    & \multicolumn{1}{c|}{\textbf{Train}} & \multicolumn{1}{c|}{\textbf{Valid}} & \textbf{Test} &                                                                                        &                                   \\ \Xhline{1.1pt}
\textbf{\bgp}                  & 52                                 & \multicolumn{1}{c|}{180}               & \multicolumn{1}{c|}{60}                  & 19,103           & 2                                                         &  $60.5\rlap{\raisebox{0.2ex}{*}}$                              \\
\textbf{\tor}                  & 175                                & \multicolumn{1}{c|}{8,115}             & \multicolumn{1}{c|}{2,706}               & 4,420            & 2                                                         & $43.9\rlap{\raisebox{0.2ex}{*}}$                                \\
\textbf{\iot}                  & 273                                & \multicolumn{1}{c|}{43,866}            & \multicolumn{1}{c|}{2,309}               & 13,128           & 2                                                                                      & 74.16                              \\
\textbf{\cookie}               & 1,689                              & \multicolumn{1}{c|}{21,046}            & \multicolumn{1}{c|}{1,108}               & 55,381           & 4                                                                                      & 71.9                              \\
\textbf{\microsoft}            & 1,804                              & \multicolumn{1}{c|}{434}               & \multicolumn{1}{c|}{8,260}               & 2,174            & 9                                                                                      & 84                                \\
\textbf{\BODMAS}               & 2,831                              & \multicolumn{1}{c|}{752}             & \multicolumn{1}{c|}{190}                 & Table~\ref{tab:bodmas-test-samples.}           & 5                                                                                     & Fig~\ref{fig:bodmas-test}                                \\
\textbf{\nprintml}             & 4,169                              & \multicolumn{1}{c|}{8,271}             & \multicolumn{1}{c|}{436}                 & 3,732            & 13                                                                                     & 75.5                             
\end{tabular}
\caption{Performance and data statistics for the 7 security tasks. F-score is reported after data normalization. * indicates F-score for only malicious class.
Test set statistics and macro F-scores across 12 \bodmas months are presented in Table~\ref{tab:bodmas-test-samples.} (Appendix) and Figure~\ref{fig:bodmas-test} respectively.}
\label{tab:table2_stats}
\end{table}

\para{\iot~\cite{tekiner2022lightweight}} A binary classification task to distinguish between cryptojacking malware and benign applications using features extracted from IoT device network traffic streams.
Each data sample comprises 273 tabular features extracted from time-series using tsfresh~\cite{christ2018time}. 
The training set contains 1,116 malicious and 42,750 benign samples. Four classifiers \ie a Logistic Regression, Gaussian Naive Bayes, Support Vector Machine and K-Nearest Neighbors are trained to report an average performance. We focus on the Logistic Regression model for simplicity, which achieves a macro F-score of 74.16\%. This task also suffers from the limited or biased training data challenges (a)-(c) (Section~\ref{sec:data-challenges-limited}).

\para{\cookie~\cite{bollinger2022automating}} A multi-class classification task which classifies website cookies into 4 privacy invasion categories (as per GDPR regulations). Each sample comprises 1,689 features that characterize several cookie properties such as name, domain, path, expiration timestamp, and flags such as 'HttpOnly' and 'HostOnly'. The original dataset contains 89k, 70k, 46k, and 15k samples across the classes, for which the authors emphasize the high human effort in labeling. To reduce computational effort, we use a random sub-sample of 8.4k, 6.6k, 4.4k, and 1.4k samples per class. An XGBoost classifier is used yielding a macro F-score of 71.9\%. This task suffers from limited or biased training data challenges (a), (d) (Section~\ref{sec:data-challenges-limited}).

\para{\microsoft~\cite{ahmadi2016novel}} A multi-class classification problem, where malware files are classified into one of 9 classes, originally from a 2015 Kaggle competition~\cite{kaggle_malware_classification}. As per the winning team's approach~\cite{ahmadi2016novel}, each data sample comprises 1,804 features representing the outputs of a hex-dump and disassembled files. Ahmadi et al.~\cite{ahmadi2016novel} achieve a 98\% macro F-score with an XGBoost classifier using the entire training set of 10,868 samples, leaving minimal room for improvement. We induce a data-challenged setting by using only \textasciitilde 4\% of the dataset (434 samples) for training. An XGBoost classifier trained on the updated dataset achieves a lower macro F-score of 84\%. The task suffers from limited or biased training data challenges (a), (b) and (d) (Section~\ref{sec:data-challenges-limited}).

\begin{figure} [h!]
\centering
\includegraphics[scale=0.8]{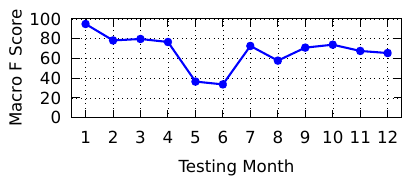}
\caption{Macro F-score for \bodmas classifier.}
\Description{A bar chart showing the Macro F-score performance of the BODMAS classifier across different training settings.}
\label{fig:bodmas-test}
\end{figure}

\para{\BODMAS~\cite{yang2021bodmas}}A  multi-class classification problem where port- able executable (PE) files collected over a period of 13 months are classified into one of 40 malware classes. Each sample comprises of 2,831 statistical features extracted using LIEF feature extractor~\cite{LIEF}. A Gradient Boosted Decision Tree (GBDT) classifier is trained on the first month's data and tested on the next 12 months. We observe that only 5 of the 40 classes show significant performance degradation. Thus, our analysis focuses on these 5 classes. The updated training set includes 49 to 367 samples per class, totaling 752 samples. As shown in Figure~\ref{fig:bodmas-test}, performance drops notably in months 5 and 6. This task suffers from limited or biased training data challenges (a), (b), and (d), as well as concept drift (Section~\ref{sec:data-challenges-limited}).

\para{\nprintml~\cite{holland2021new, cicids2017}} A multi-class classification of network traffic originating from one of 13 operating systems (OS) for intrusion detection. Using the nPrintML framework~\cite{holland2021new}, raw traffic is converted into tabular form, and AutoML~\cite{he2021automl} suggests a classifier. Each sample includes 4,169 binary features indicating the presence or absence of protocol fields and payloads. The training set contains between 289 and 689 samples per class, totaling 8,271 samples. An AutoML-selected Light Gradient Boosted Machine (LGBM) classifier is trained, achieving a macro F-score of 75.5\%. This task faces limited or biased training data challenges (a), (b), and (d) (Section~\ref{sec:data-challenges-limited}).

\section{Data Augmentation Techniques}
\label{sec:data_aug}
\noindent 
To perform data augmentation using synthetic data, we use state-of-the-art GenAI tools and propose a new GenAI scheme called \textbf{\sysname{}}. We also consider traditional (non-GenAI) augmentation tools. 

\subsection{Augmentation strategy}
\label{sec:augment-strategy}

\noindent All the tasks chosen for our analysis have imbalanced class distributions. We use a simple strategy for data augmentation---we balance the class distribution, \ie we generate enough synthetic samples for all the minority classes, until they match the size of the largest class. This class-balancing augmentation strategy can be implemented by generating synthetic samples in 2 ways: (1) \textit{Class-conditioning}: In this case, the data synthesis tool provides an explicit mechanism for conditioning on the class label while generating samples, or we resort to rejection sampling~\cite{bishop2006pattern}, until we obtain enough samples for each class. All existing GenAI schemes use this method.
(2) \textit{Sample-conditioning:} Sample-conditioned techniques generate synthetic data by sampling in the vicinity of an existing  sample in the training set (\ie real samples). The generated sample is assigned the same label as the conditioned sample. Our new GenAI scheme, \sysname{} can perform both sample- and class-conditioning.

\subsection{Existing GenAI tools for tabular data}
\label{sec:data_aug_genai}

\noindent We describe each GenAI scheme used in our study. The implementation details are in Section~\ref{subsec:genai-additional} (Appendix).

\para{\tvae~\cite{tvae/ctgan}}\tvae is based on a Variational Autoencoder and uses a class-conditional sampling technique. Specifically, \tvae comprises an encoder network that maps samples to a lower dimensional latent space, and a decoder network that maps back to the feature space. \tvae uses special data processing schemes to handle tabular data challenges. To handle multimodal continuous-valued columns, \tvae uses Variational Gaussian Mixture (VGM) models ~\cite{bishop2006pattern} to estimate modes and fit a Gaussian mixture.
To generate a synthetic sample, a \tvae  samples from the latent space and feeds this sample through the decoder. \tvae  does not natively support class-conditional sampling, i.e., generating synthetic data from a specific class. As such, we employ rejection sampling~\cite{bishop2006pattern} to obtain samples for any given class. We disregard comparison with another VAE-based generator~\cite{dai2019generative} as it is limited to binary classification.

\para{\ctabgan~\cite{ctabgan+}}\ctabgan is based on a Conditional Generative Adversarial Network (cGAN) and uses a class-conditional sampling technique. \ctabgan belongs to a series of GAN-based techniques proposed for synthetic tabular data generation~\cite{tvae/ctgan, gulrajani2017improved, ctabgan}. Specifically, it comprises generator and discriminator networks as part of a GAN~\cite{goodfellow2014generative}. Akin to image generation, \ctabgan uses Wasserstein distance with gradient penalty as loss~\cite{gulrajani2017improved} term to improve training stability for underrepresented sample patterns. Also, it uses \textit{training-by-sampling} method to over-sample rare data points in the training batches for their emphasized learning.
To generate a synthetic sample, \ctabgan randomly samples from a Gaussian distribution, and feeds this sample through the generator. Continuous features are processed using the same approach as used by \tvae, \ie mode-specific normalization~\cite{tvae/ctgan}. Unlike \tvae, \ctabgan supports class-conditional sampling using a class conditional vector to guide the generator. 
We use \ctabgan because it outperforms other GAN-based tabular data generators~\cite{tvae/ctgan, gulrajani2017improved, ctabgan}. 

\para{\tabddpm~\cite{tabddpm}}\tabddpm leverages a Diffusion model and uses a class-conditional sampling technique. It comprises a network that progressively denoises Gaussian-noised samples as part of a reverse Markov process. For generation, it samples randomly from a Gaussian distribution and recursively feeds this sample through the network. 
\tabddpm performs hyperparameter search for optimal architecture parameters.

\para{\tabsyn~\cite{tabsyn}}\tabsyn leverages a VAE in conjunction with a Diffusion model operating in the VAE latent space and uses a class-conditional sampling technique. To generate a synthetic sample, \tabsyn draws from a Gaussian distribution and passes it through a Diffusion model to obtain a VAE latent space sample, which is then decoded using the VAE's decoder. \tabsyn does not natively support class-conditional sampling. Like TVAE, we use rejection-sampling to obtain samples for the required class. \tabsyn outperforms other Diffusion-based tabular data generators, \ie CoDi~\cite{lee2023codi} and STaSy~\cite{kim2022stasy}.

\para{\great~\cite{GreaT}}\great leverages an LLM for tabular data generation. Specifically, \great encodes tabular features to natural language text representation for autoregressive language modeling with an LLM. During training, a pretrained LLM (distilGPT-2~\cite{sanh2019distilbert}) is fine-tuned on the encoded textual data. To generate a synthetic sample, \great  encodes feature names or values and autoregressively samples the remainder values as text, then decodes it back to the tabular space. In theory, \great can generate synthetic samples by conditioning on any field of the feature vector. However, it does not propose a method for targeted conditioning—\ie generating samples in the vicinity of  certain samples. Thus, we condition only on the class field for class-conditioning, and treat \great as a class-conditional approach in our evaluation.

\para{\realtabformer~\cite{solatorio2023realtabformer}} \realtabformer (like \great) fine-tunes a pre-trained LLM (e.g., GPT-2) to generate realistic tabular data. Continuous features are encoded as fixed-length, sub-columnar string tokens mapped to a unique vocabulary. Training employs early stopping based on similarity between a held-out validation set and synthetic samples to prevent overfitting. During generation, output is constrained to generate tokens belonging to the tokens in the training vocabulary, and class-conditional sampling enables controlled data generation. Other LLM-based generators were not considered because CLLM~\cite{seedatcurated} is proposed for datasets under 100 samples, and TabMT~\cite{gulati2024tabmt} did not make their code available.

\subsection{Traditional data augmentation tools}
\label{subsec:traditional-baselines}
\noindent
We select two off-the-shelf non-GenAI baseline methods for comparison. These ``traditional'' approaches use rule-based techniques to weight existing samples and augment data under specific constraints.
Implementation details are provided in Section~\ref{subsec:genai-additional}.

\para{\smote~\cite{smote}}\smote is a popular sample-conditioned technique that leverages linear interpolation between samples. Specifically, given an existing data sample from the real dataset, it generates synthetic data by sampling along the lines that connect the sample to its $k$-nearest neighbors. \smote does not scale well to high-dimensional imbalanced datasets~\cite{blagus2013smote}. A plethora of methods build on \smote~\cite{fernandez2018smote}. One of them, DeepSMOTE~\cite{dablain2022deepsmote} uses a latent-space interpolation via an encoder-decoder, but supports only image data, making it unsuitable for tabular tasks (Appendix~\ref{sub-sec:deepsmote}).
Despite its limitations, we include \smote for comparison, as it remains the most widely used data augmentation method~\cite{fernandez2018smote}.

\para{\mcccr~\cite{mc-ccr}}\mcccr is a sample-conditioned technique with specific adaptations for the multi-class setting. Before generating synthetic data, \mcccr performs a ``cleaning'' stage, where samples with ``wrong'' labels are repositioned by moving samples from other classes which are close to samples of the target class. Re-positioning is achieved by translating the mislabeled samples to the surface of a sphere around it. 
If many wrong labeled samples are re-positioned around a data sample, it is considered ``difficult''. \mcccr thus generates synthetic data by sampling within the sphere, with more synthetic data being generated for ``difficult'' samples.
Additional details are in Appendix~\ref{sub-sec:mcccr-details}. 

\para{Other methods} We do not evaluate other SMOTE-based oversampling methods, as MC-CCR has been shown to outperform both other multi-class oversampling techniques and simple GAN-based generators in highly imbalanced settings~\cite{dablain2022deepsmote}. We also exclude MC-RBO~\cite{krawczyk2019radial}, since unlike MC-CCR, it lacks a cleaning strategy to remove outliers before augmentation in multi-class scenarios.

\begin{figure}[t!]
    \centering
    \includegraphics[width=0.8\columnwidth]{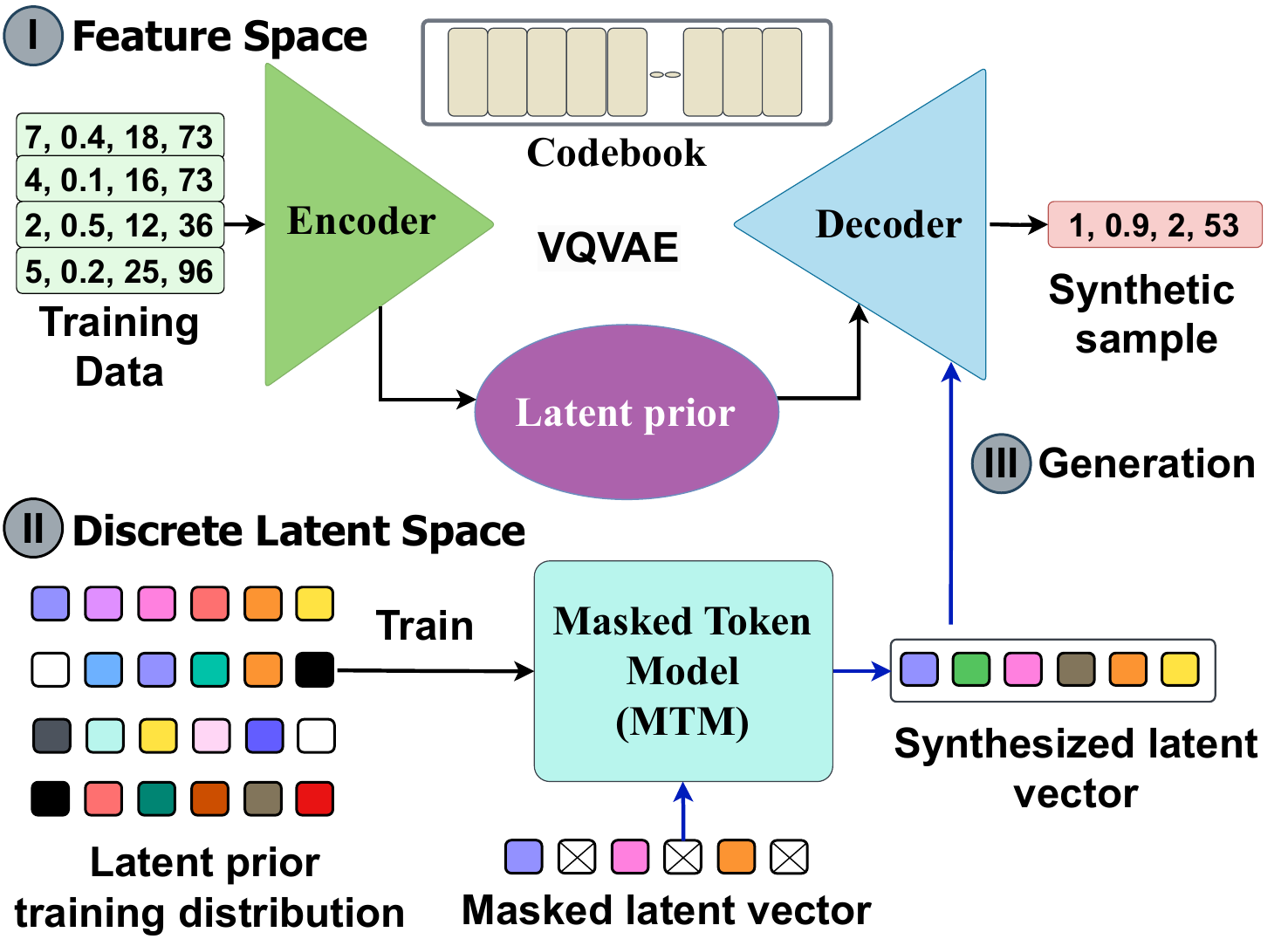}
    \caption{Architecture of \nimai framework.}
 \Description{A diagram showing the architecture of the Nimai framework, including components for data generation, model training, and evaluation, connected in a sequential pipeline.}
    \label{fig:nimai-framwork}
\end{figure}

\subsection{\nimai: Enabling controlled data synthesis} 
\label{subsec:nimai-mainframe}
\noindent
Recall that existing GenAI schemes are class-conditional synthesis schemes. 
To alleviate complex biases within a class, the defender may require more control over the data generation process. Class-conditional synthesis or uncontrolled synthesis that creates many synthetic samples to broadly mimic real samples may not be the best strategy in all cases.
We propose a new GenAI scheme called \sysname{} that provides highly controlled data synthesis through a sample-conditioning scheme. Such an approach would help target regions of the data manifold that are underrepresented. That said, \sysname{} also supports class-conditional synthesis. In addition, \sysname{} provides automated mechanisms to efficiently tune its hyper-parameters.
Recall that among the existing GenAI schemes, only \tabddpm offers strategies for hyper-parameter tuning when fitting the model to diverse datasets.
\textit{We emphasize that the goal of \sysname{} is not to provide a ``one-fits-all'' solution that supersedes the GenAI schemes discussed in Section~\ref{sec:data_aug_genai}; rather, we aim to understand the impact of using sample-conditioning, compared to a class-conditional approach, for security tasks.}

\para{Architecture and training} \sysname{} is a VAE-based GenAI model. \sysname{} adapts a Vector Quantized Variational AutoEncoder (VQ-VAE)~\cite{vqvae} to enable a discrete latent space. We convert the 2D discrete latent tokens used for images into 1D latent vectors suitable for tabular data. A key innovation lies in integrating a BERT-based masked-token-model (MTM) in the latent space, enabling sample-based generation with controllable variation via a masking ratio and supporting class-conditional generation.
Unlike GANs, VAEs are easier to train due to their tractable likelihood loss~\cite{chang2022maskgit}, and their likelihood maximization encourages greater sample diversity~\cite{van2016conditional}. Compared to Diffusion models, VAEs offer faster sampling, avoiding the costly iterative denoising process. 
We aim to create synthetic samples in the ``neighborhood'' of existing samples (\eg in minority classes), thus improving generalization performance. \textit{Our key idea is to use a \textbf{discrete latent space} for the VAE model to controllably produce synthetic samples that generalize better}. In comparison, existing VAE-based methods like \tvae and \tabsyn encode features into a continuous latent space, which can result in prior holes and posterior collapse~\cite{aneja2021contrastive, lucas2019understanding} impacting performance.
Recently, discrete latent codes have driven success across tasks, especially in text generation~\cite{keskar2019ctrl}, as well as image and video generation~\cite{bai2024sequential}. For instance, LLMs model distributions over discrete variables, enabling highly controlled text generation~\cite{keskar2019ctrl, dathathri2019plug}. 

Figure~\ref{fig:nimai-framwork} depicts \sysname{}'s architecture. We implement a discrete latent space via a shared embedding \textit{codebook} with $K$ vectors (one per categorical value). Transformer blocks serve as encoder/decoder, and tabular data is normalized to [0,1] for stable learning.
An input tabular sample, along with its class label, is first converted to a continuous representation by the encoder block.  This continuous output is discretized by identifying the nearest embedding vectors in the (trainable) codebook, \ie discretized by using the indices of the nearest vectors in the codebook. The decoder receives this discrete latent vector as input. The model is trained to bound log-likelihood via the evidence lower bound (ELBO), while also optimizing losses that align codebook vectors with encoder outputs and enforce encoder commitment.
A full description of the training objective is provided in Appendix~\ref{appendix:training-objective-nimai}.

\para{Synthesizing samples}
To synthesize generalizable samples, we need to \textit{train a prior over the discrete random variables.} This mechanism enables choosing a suitable prior to generate samples near existing ones, including underrepresented samples, enhancing targeted sample synthesis. Once trained, new discrete latent vectors can be sampled from this prior and fed to the decoder to generate new synthetic samples. 

We develop \textbf{Masked Token Modeling (MTM)} to build the prior. This is inspired by work in the image~\cite{chang2022maskgit} and natural language domain~\cite{kenton2019bert}.
More concretely, to perform sample-conditioning, we use the encoder to project a sample in the input feature space to the latent space, and then mask, \ie ``blank-out'' a subset of the latent features. 
Our MTM technique takes a feature vector and its class label as input, and predicts masked feature values using surrounding context. It is implemented using the BERT model~\cite{kenton2019bert}.
We fill-in (infer) new features in a non-autoregressive manner in a constant number of timesteps.  A detailed description of the MTM objective is provided in Appendix~\ref{appendix:mtm-objective}.
At generation time (Stage III of Figure~\ref{fig:nimai-framwork}), a latent feature vector with the ``blanks-filled-in'' (synthesized vector) with its class label is passed through the decoder, mapping it to a nearby sample in the input feature space. We ensure that synthetic data preserve the value ranges of the original feature space.

\textit{Note that sample-conditioning via MTM reduces to the class-} \textit{conditioned case when all the latent features are masked out.} As such, \sysname{} inherently allows for interpolating between sample-conditioned or class- conditioned sampling, i.e., treats the latent feature \textit{masking ratio} as a hyper-parameter that can be chosen using the validation set. \textit{We refer to the sample-conditioned and class-conditioned versions of \sysname{} as \textbf{\nbase} and \textbf{\nrandom}, respectively.} 
\sysname{} uses a hyperparameter search with the ASHA algorithm\cite{li2020system} across various architectural choices (Table~\ref{tab:hyper-parameters}, Appendix), making it adaptable to diverse security data. Details are provided in Appendix~\ref{subsec:nimai-hyper-search}.

\section{Evaluation Setup}
\label{sec:exp_setup}

\para{Feature Pre-processing}
Tabular data in some of our tasks contains mixed data types like continuous and discrete-valued features. Training generative models on mixed feature types is a non-trivial task~\cite{ctabgan}. We leave this aspect for future work. This work focuses on alleviating the data challenges mentioned in Section~\ref{sec:data-challenges-limited}. In our evaluation, we standardize all feature values between 0 and 1, irrespective of the data type. 
All features are normalized to [0,1] with a MinMax scaler (using clipping to keep held-out values in range). Normalized real data trains GenAI schemes, and the same scaler ensures that all synthetic data also lies within [0,1].
We find that this scaling has minimal impact on model performance in 6/7 datasets where the end defense classifier is a tree-based model \eg XGBoost, LightGBM. The exception was \iot, where a non-tree based classifier, \ie Logistic Regression is used, the performance increases from 46\% to 74.16\% macro F-score after data standardization. Note that despite the improvement, all augmentation schemes (GenAI and traditional approaches) are evaluated on the same standardized dataset, \ie relative performance comparison. 

\para{Performance metrics}
For each task in Section~\ref{sec:tasks-metrics}, we apply the same training setup (classifier and hyperparameters) with and without augmentation.
We measure the \textit{macro average F-score} for all tasks, except for \tor and \bgp, where we report the F-score for the malicious class (see Section~\ref{sec:tasks-metrics}).
For any given $N-$class classifier, let $P_i$ and $R_i$ represent the Precision and Recall on the $i^{\text{th}}$ class. The macro average F-score is then given by:

\begin{equation}
 F = \dfrac{2}{N}\sum_{i=1}^{N}\frac{ P_{i} \cdot R_i}{P_{i} + R_{i}} 
\end{equation}
which may be viewed as an unweighted average of individual F-scores across each of the $N$ classes. We focus on evaluating \textit{relative} gains in classifier performance. Gain may also be negative if there is performance degradation after data augmentation. Let $F_{real}$ and $F_{aug}$ represent the F-scores of the classifier before and after data augmentation, respectively. We then measure relative gain $\Delta G$ as:
\begin{equation}
    \Delta G = \frac{F_{aug} - F_{real}} {F_{real}}
\end{equation}
When measuring $\Delta G$, we repeat experiments across 10 trials with different random seeds and compute the mean and standard deviation of F-scores. Note that for two tasks \bodmas and \microsoft, when computing $\Delta G$, we omit classes with too few test set samples ($\leq$ 20). Higher, positive values of $\Delta G$ indicate better performance from data augmentation.
Our setup enables direct attribution of performance changes (\ie $\Delta G$) to the synthetic data used for augmentation and supports fair comparison across GenAI and non-GenAI schemes. For each task, all augmentation methods use the same classifier configuration, differing only in the augmented datasets.

\para{Identifying unreliable measurements} 
We use the Coefficient of Variation (CV)~\cite{abdi2010coefficient} to identify unreliable $\Delta G$ measurements, computed as the ratio of standard deviation to mean. When CV exceeds 1, we deem the augmentation scheme to have high dispersion and the measurement unreliable. 
 We focus our discussions primarily on reliable schemes and label unreliable schemes as such.

\para{Statistical tests}
In certain cases, comparing reliable data augmentation schemes using a simple ordering of their mean $\Delta G$ scores can be problematic, \eg when a scheme with larger mean is accompanied by a larger standard deviation. To resolve these ambiguities, we compare $\Delta G$ rankings using the non-parametric Kruskal–Wallis test~\cite{kruskal1952use}, which does not assume homogeneity of variance~\cite{montgomery2019design}. When Kruskal-Wallis test ranks differ, we apply Dunn's~\cite{dunn1964multiple} test for post-hoc pairwise ranking comparisons.

\section{Evaluation: Pre-Deployment}
\label{sub-sec:genai-eval}
\noindent The focus here is on improving performance via data augmentation \textit{before deploying} the classifier (see Figure~\ref{fig:nimai-overview}(a)). Evaluation using the 7 data-challenged tasks (Section~\ref{sec:measurement-study}) are divided into two parts: (1) cases where GenAI schemes demonstrate clear potential, and (2) cases where most GenAI schemes fall short, \ie failed to significantly boost performance. 

We list the quantity of synthetic samples added per task in Table~\ref{tab:synth-samples-added} (Appendix) and the training times for all  GenAI schemes in Table~\ref{tab:genai-train-times} (Appendix).

\begin{table}[t!]
\centering
\small
\setlength{\tabcolsep}{2pt}
\setlength\extrarowheight{1.5pt}
\begin{tabular}{c"cccccc}
                                     & \multicolumn{6}{c}{\textbf{Mean $\Delta G$ with standard deviation}}                                                                                                                                                                                                                                    \\ \cline{2-7} 
\multirow{-2}{*}{\textbf{Technique}} & \multicolumn{2}{c|}{\textbf{\iot}}                                                           & \multicolumn{2}{c|}{\textbf{\bgp}}                                                             & \multicolumn{2}{c}{\textbf{\microsoft}}                                          \\ \Xhline{1.1pt}

\textbf{\nbase}                      & \multicolumn{1}{c|}{17.37 (0.28)}  & \multicolumn{1}{c|}{\cellcolor[HTML]{55a630}$\uparrow$} & \multicolumn{1}{c|}{32.61 (3.76)}  & \multicolumn{1}{c|}{\cellcolor[HTML]{55a630}$\uparrow$} & \multicolumn{1}{c|}{6.89 (6.47)}  & {\cellcolor[HTML]{55a630}$\uparrow$}\\ \hline
\textbf{\nrandom}                    & \multicolumn{1}{c|}{3.93 (0.68)}   & \multicolumn{1}{c|}{\cellcolor[HTML]{55a630}$\uparrow$} & \multicolumn{1}{c|}{27.65 (0.81)}  & \multicolumn{1}{c|}{\cellcolor[HTML]{55a630}$\uparrow$} & \multicolumn{1}{c|}{11.67 (5.15)} & {\cellcolor[HTML]{55a630}$\uparrow$} \\ \hline
\textbf{\tvae}                       & \multicolumn{1}{c|}{2.02 (0.66)}   & \multicolumn{1}{c|}{\cellcolor[HTML]{55a630}$\uparrow$} & \multicolumn{1}{c|}{-20.83 (3.22)} & \multicolumn{1}{c|}{\cellcolor[HTML]{da5552}$\downarrow$} & \multicolumn{1}{c|}{11.46 (5.88)} & {\cellcolor[HTML]{55a630}$\uparrow$} \\ \hline
\textbf{\tabsyn}                     & \multicolumn{1}{c|}{-0.69 (1.21)}  & \multicolumn{1}{c|}{\cellcolor[HTML]{CCCCCC}-}          & \multicolumn{1}{c|}{-2.78 (5.94)}  & \multicolumn{1}{c|}{\cellcolor[HTML]{CCCCCC}-}          & \multicolumn{1}{c|}{$\times$}     & {\cellcolor[HTML]{CCCCCC}$\times$} \\ \hline
\textbf{\great}                      & \multicolumn{1}{c|}{$\times$}      & \multicolumn{1}{c|}{\cellcolor[HTML]{CCCCCC}$\times$}   & \multicolumn{1}{c|}{-71.7 (2.56)}  & \multicolumn{1}{c|}{\cellcolor[HTML]{da5552}$\downarrow$} & \multicolumn{1}{c|}{$\times$}       & \cellcolor[HTML]{CCCCCC}$\times$     \\ \hline
\textbf{\realtabformer}             & \multicolumn{1}{c|}{$\times$}   & \multicolumn{1}{c|}{\cellcolor[HTML]{CCCCCC}$\times$ } & \multicolumn{1}{c|}{-15.35 (5.37)}  & \multicolumn{1}{c|}{\cellcolor[HTML]{da5552}$\downarrow$} & \multicolumn{1}{c|}{$\times$}  & {\cellcolor[HTML]{CCCCCC}$\times$ } \\ \hline
\textbf{\ctabgan}                    & \multicolumn{1}{c|}{5.84 (1.18)}   & \multicolumn{1}{c|}{\cellcolor[HTML]{55a630}$\uparrow$} & \multicolumn{1}{c|}{14.83 (4.07)}  & \multicolumn{1}{c|}{\cellcolor[HTML]{55a630}$\uparrow$} & \multicolumn{1}{c|}{13.53 (3.92)} & {\cellcolor[HTML]{55a630}$\uparrow$}           \\ \hline
\textbf{\tabddpm}                    & \multicolumn{1}{c|}{-38.01 (0.00)} & \multicolumn{1}{c|}{\cellcolor[HTML]{CCCCCC}$\times$}   & \multicolumn{1}{c|}{0.51 (6.76)}   & \multicolumn{1}{c|}{\cellcolor[HTML]{CCCCCC}-}          & \multicolumn{1}{c|}{14.34 (0.37)} & {\cellcolor[HTML]{55a630}$\uparrow$} \\ \hline
\textbf{\smote}                      & \multicolumn{1}{c|}{14.23 (0.23)}  & \multicolumn{1}{c|}{\cellcolor[HTML]{55a630}$\uparrow$} & \multicolumn{1}{c|}{12.63 (4.56)}  & \multicolumn{1}{c|}{\cellcolor[HTML]{55a630}$\uparrow$} & \multicolumn{1}{c|}{-0.22 (0.38)} & {\cellcolor[HTML]{CCCCCC}\textbf{-}}  \\ \hline
\textbf{\mcccr}                      & \multicolumn{1}{c|}{0.27 (0.37)}   & \multicolumn{1}{c|}{\cellcolor[HTML]{CCCCCC}-}          & \multicolumn{1}{c|}{-8.64 (0.66)}  & \multicolumn{1}{c|}{\cellcolor[HTML]{da5552}$\downarrow$} & \multicolumn{1}{c|}{-0.15 (0.18)} & {\cellcolor[HTML]{CCCCCC}-} \\

\end{tabular}
\caption{Mean $\Delta G$ with standard deviation for \iot, \bgp, and \microsoft tasks, where GenAI shows potential. $\uparrow$ and $\downarrow$ indicate positive and negative gains, respectively. `-' indicates unreliable gains (CV $\geq$ 1), and `$\times$' indicates model failures.}
\label{tab:genai-success-results}
\end{table}

\begin{table*}[ht!]
\centering

 \small
\setlength{\tabcolsep}{2pt}
\setlength\extrarowheight{1.5pt}
\begin{tabular}{c"cccccccccccccccc}
                                     & \multicolumn{16}{c}{\textbf{Mean $\Delta G$ with standard deviation for months}}                                                                                                                                                                                                                                                                                                                                                                                                                    \\ \cline{2-17} 
\multirow{-2}{*}{\textbf{Technique}} & \multicolumn{2}{c|}{\textbf{2}}                                                              & \multicolumn{2}{c|}{\textbf{3}}  & \multicolumn{2}{c|}{\textbf{5}}                                                              & \multicolumn{2}{c|}{\textbf{6}}                                                              & \multicolumn{2}{c|}{\textbf{7}}                                                              & \multicolumn{2}{c|}{\textbf{9}}                                                                & \multicolumn{2}{c|}{\textbf{10}}      & \multicolumn{2}{c}{\textbf{11}}                                     \\ \Xhline{1.1pt}
\textbf{\nbase}      & \multicolumn{1}{c|}{2.43 (7.75)}    & \multicolumn{1}{c|}{\cellcolor[HTML]{CCCCCC}-}            & \multicolumn{1}{c|}{-3.64 (5.42)}  & \multicolumn{1}{c|}{\cellcolor[HTML]{CCCCCC}-}                      & \multicolumn{1}{c|}{36.68 (21.60)} & \multicolumn{1}{c|}{\cellcolor[HTML]{55A630}$\uparrow$} & \multicolumn{1}{c|}{30.45 (15.69)} & \multicolumn{1}{c|}{\cellcolor[HTML]{55A630}$\uparrow$} & \multicolumn{1}{c|}{2.10 (4.41)}   & \multicolumn{1}{c|}{\cellcolor[HTML]{CCCCCC}-}          & \multicolumn{1}{c|}{8.00 (12.66)}  & \multicolumn{1}{c|}{\cellcolor[HTML]{CCCCCC}-}            & \multicolumn{1}{c|}{6.84 (5.27)}    & \multicolumn{1}{c|}{\cellcolor[HTML]{55A630}$\uparrow$}  & \multicolumn{1}{c|}{4.03 (11.49)}   & \cellcolor[HTML]{CCCCCC}- \\ \hline
\textbf{\nrandom}      & \multicolumn{1}{c|}{6.88 (3.16)}    & \multicolumn{1}{c|}{\cellcolor[HTML]{55A630}$\uparrow$}   & \multicolumn{1}{c|}{5.57 (4.78)}  &     \multicolumn{1}{c|}{\cellcolor[HTML]{55A630}$\uparrow$}   &          \multicolumn{1}{c|}{59.42 (8.69)}  & \multicolumn{1}{c|}{\cellcolor[HTML]{55A630}$\uparrow$} & \multicolumn{1}{c|}{28.24 (21.25)} & \multicolumn{1}{c|}{\cellcolor[HTML]{55A630}$\uparrow$} & \multicolumn{1}{c|}{4.34 (3.47)}   & \multicolumn{1}{c|}{\cellcolor[HTML]{55A630}$\uparrow$} & \multicolumn{1}{c|}{21.73 (11.32)} & \multicolumn{1}{c|}{\cellcolor[HTML]{55A630}$\uparrow$}   & \multicolumn{1}{c|}{10.23 (2.60)}   & \multicolumn{1}{c|}{\cellcolor[HTML]{55A630}$\uparrow$} & \multicolumn{1}{c|}{12.77 (2.01)}   & \cellcolor[HTML]{55A630}$\uparrow$ \\ \hline
\textbf{\tvae}         & \multicolumn{1}{c|}{-10.05 (10.01)} & \multicolumn{1}{c|}{\cellcolor[HTML]{DA5552}$\downarrow$} & \multicolumn{1}{c|}{-8.81 (10.34)}      & \multicolumn{1}{c|}{\cellcolor[HTML]{CCCCCC}-}         & \multicolumn{1}{c|}{52.09 (18.00)} & \multicolumn{1}{c|}{\cellcolor[HTML]{55A630}$\uparrow$} & \multicolumn{1}{c|}{50.06 (20.81)} & \multicolumn{1}{c|}{\cellcolor[HTML]{55A630}$\uparrow$} & \multicolumn{1}{c|}{-7.71 (12.43)} & \multicolumn{1}{c|}{\cellcolor[HTML]{CCCCCC}-}          & \multicolumn{1}{c|}{19.65 (9.06)}  & \multicolumn{1}{c|}{\cellcolor[HTML]{55A630}$\uparrow$}   & \multicolumn{1}{c|}{-4.70 (15.67)}  & \multicolumn{1}{c|}{\cellcolor[HTML]{CCCCCC}-}      & \multicolumn{1}{c|}{-16.64 (17.09)} & \cellcolor[HTML]{CCCCCC}-       \\ \hline
\textbf{\tabsyn}       & \multicolumn{1}{c|}{-11.10 (5.23)}  & \multicolumn{1}{c|}{\cellcolor[HTML]{DA5552}$\downarrow$} & \multicolumn{1}{c|}{-16.77 (5.67)}       & \multicolumn{1}{c|}{\cellcolor[HTML]{DA5552}$\downarrow$}        & \multicolumn{1}{c|}{15.77 (10.86)} & \multicolumn{1}{c|}{\cellcolor[HTML]{55A630}$\uparrow$} & \multicolumn{1}{c|}{13.84 (12.73)} & \multicolumn{1}{c|}{\cellcolor[HTML]{55A630}$\uparrow$} & \multicolumn{1}{c|}{-7.09 (9.86)}  & \multicolumn{1}{c|}{\cellcolor[HTML]{CCCCCC}-}          & \multicolumn{1}{c|}{-14.49 (8.49)} & \multicolumn{1}{c|}{\cellcolor[HTML]{DA5552}$\downarrow$} & \multicolumn{1}{c|}{-13.20 (15.98)} & \multicolumn{1}{c|}{\cellcolor[HTML]{CCCCCC}-}    & \multicolumn{1}{c|}{-20.98 (25.20)} & \cellcolor[HTML]{CCCCCC}-        \\ \hline
\textbf{\ctabgan}        & \multicolumn{1}{c|}{1.98 (6.72)}    & \multicolumn{1}{c|}{\cellcolor[HTML]{CCCCCC}-}            & \multicolumn{1}{c|}{0.39 (7.37)}      & \multicolumn{1}{c|}{\cellcolor[HTML]{CCCCCC}-}            & \multicolumn{1}{c|}{55.99 (21.87)} & \multicolumn{1}{c|}{\cellcolor[HTML]{55A630}$\uparrow$} & \multicolumn{1}{c|}{45.45 (33.85)} & \multicolumn{1}{c|}{\cellcolor[HTML]{55A630}$\uparrow$} & \multicolumn{1}{c|}{1.59 (1.64)}   & \multicolumn{1}{c|}{\cellcolor[HTML]{CCCCCC}-}          & \multicolumn{1}{c|}{14.09 (15.36)} & \multicolumn{1}{c|}{\cellcolor[HTML]{CCCCCC}-}            & \multicolumn{1}{c|}{3.70 (4.20)}    & \multicolumn{1}{c|}{\cellcolor[HTML]{CCCCCC}-}  & \multicolumn{1}{c|}{-5.29 (3.45)}   & \cellcolor[HTML]{DA5552}$\downarrow$        \\ \hline
\textbf{\tabddpm}          & \multicolumn{1}{c|}{-2.02 (2.46)}   & \multicolumn{1}{c|}{\cellcolor[HTML]{CCCCCC}-}            & \multicolumn{1}{c|}{-2.71 (2.09)}   & \multicolumn{1}{c|}{\cellcolor[HTML]{DA5552}$\downarrow$}         & \multicolumn{1}{c|}{27.20 (9.81)}  & \multicolumn{1}{c|}{\cellcolor[HTML]{55A630}$\uparrow$} & \multicolumn{1}{c|}{18.96 (3.77)}  & \multicolumn{1}{c|}{\cellcolor[HTML]{55A630}$\uparrow$} & \multicolumn{1}{c|}{2.57 (0.45)}   & \multicolumn{1}{c|}{\cellcolor[HTML]{55A630}$\uparrow$} & \multicolumn{1}{c|}{-0.63 (1.05)}  & \multicolumn{1}{c|}{\cellcolor[HTML]{CCCCCC}-}            & \multicolumn{1}{c|}{10.45 (0.42)}   & \multicolumn{1}{c|}{\cellcolor[HTML]{55A630}$\uparrow$}  & \multicolumn{1}{c|}{0.03 (0.37)}    & \cellcolor[HTML]{CCCCCC}-  \\ \hline
\textbf{\smote}        & \multicolumn{1}{c|}{-3.38 (1.44)}   & \multicolumn{1}{c|}{\cellcolor[HTML]{DA5552}$\downarrow$} & \multicolumn{1}{c|}{-3.00 (2.69)}     & \multicolumn{1}{c|}{\cellcolor[HTML]{DA5552}$\downarrow$}           & \multicolumn{1}{c|}{10.96 (11.77)} & \multicolumn{1}{c|}{\cellcolor[HTML]{CCCCCC}-}          & \multicolumn{1}{c|}{5.45 (6.72)}   & \multicolumn{1}{c|}{\cellcolor[HTML]{CCCCCC}-}          & \multicolumn{1}{c|}{1.78 (0.96)}   & \multicolumn{1}{c|}{\cellcolor[HTML]{55A630}$\uparrow$} & \multicolumn{1}{c|}{0.45 (1.43)}   & \multicolumn{1}{c|}{\cellcolor[HTML]{CCCCCC}-}            & \multicolumn{1}{c|}{3.27 (4.29)}    & \multicolumn{1}{c|}{\cellcolor[HTML]{CCCCCC}-}    & \multicolumn{1}{c|}{0.13 (0.19)}    & \cellcolor[HTML]{CCCCCC}-        \\ \hline
\textbf{\mcccr}          & \multicolumn{1}{c|}{-2.14 (5.79)}   & \multicolumn{1}{c|}{\cellcolor[HTML]{CCCCCC}-}            & \multicolumn{1}{c|}{-2.55 (2.03)}     & \multicolumn{1}{c|}{\cellcolor[HTML]{DA5552}$\downarrow$}         & \multicolumn{1}{c|}{-0.05 (9.91)}  & \multicolumn{1}{c|}{\cellcolor[HTML]{CCCCCC}-}          & \multicolumn{1}{c|}{1.70 (5.56)}   & \multicolumn{1}{c|}{\cellcolor[HTML]{CCCCCC}-}          & \multicolumn{1}{c|}{-1.39 (5.35)}  & \multicolumn{1}{c|}{\cellcolor[HTML]{CCCCCC}-}          & \multicolumn{1}{c|}{0.72 (1.49)}   & \multicolumn{1}{c|}{\cellcolor[HTML]{CCCCCC}-}            & \multicolumn{1}{c|}{-4.55 (9.98)}   & \multicolumn{1}{c|}{\cellcolor[HTML]{CCCCCC}-}   & \multicolumn{1}{c|}{-7.82 (15.93)}  & \cellcolor[HTML]{CCCCCC}-        \\ 

\end{tabular}
\caption{
Mean $\Delta G$ with standard deviation for \bodmas task for 2, 3, 5, 6, 7, 9, 10 \& 11 months. $\uparrow$ and $\downarrow$ indicate positive and negative gains, respectively. `-' indicates unreliable gains (CV $\geq$ 1). \great and \realtabformer's failures are not shown here. Results for months 1, 4, 8 \& 12  are in Table~\ref{tab:bodmas-noeval-months} (Appendix) where no gains were observed.}
\label{tab:bodmas-months2356791011}
\end{table*}

\subsection{Cases where GenAI shows potential}
\label{subsubsec:genai-potential}
\noindent We evaluate \iot, \bgp, \microsoft and \bodmas tasks. Recall that this covers two broad categories of tasks: \iot and \bgp classifiers are based on network features, while \microsoft and \bodmas classifiers are based on features extracted from the software binary.

\para{\iot} The \iot task suffers from limited or biased data challenges (a)-(c) (Section~\ref{sec:data-challenges-limited}), with a notable challenge of high class imbalance ratio of 38:1 (Section~\ref{sec:tasks-metrics}). Recall the task has a total of 43,866 training samples of which only 1,116 are attack samples. Results are in Table~\ref{tab:genai-success-results}.

\noindent
\textsl{{\para{Finding 1}\textit{Sample-conditioned GenAI schemes have the potential to boost performance in highly imbalanced class settings in binary classification tasks, outperforming traditional data augmentation schemes.}}}
We observe that the best-performing approaches use sample-conditioning. \nbase achieves a mean $\Delta G$ of 17.37\% outperforming all the schemes. Class-conditional GenAI approaches also demonstrate positive $\Delta G$, notably \tvae, \ctabgan, and \nrandom. One possible reason for \ctabgan's improved performance over \nrandom and \tvae is due to its special loss terms designed to handle imbalanced data. \nrandom achieves a mean $\Delta G$ of 3.93\%, surpassing \tvae (2.02\%), likely due to the quantized space of \sysname{}. While \smote outperforms most GenAI schemes, it falls short of \nbase, underscoring GenAI’s potential to advance augmentation. In contrast, \tabsyn and \mcccr show unreliable behavior.

\noindent
\textit{Failure cases.} Only 3/6 existing GenAI tools (\ie excluding \sysname{}) were successfully instantiated on the \iot task data.
We find that \tabddpm, \great, and \realtabformer fail to produce usable classifiers. In particular, \great generates an excessively large language representation of the 273 \iot features, requiring about 5,714 hours for a single epoch on an NVIDIA A100 GPU (See Table~\ref{tab:genai-train-times} (Appendix)). {\realtabformer fails to train on \iot as it does not reach stopping criteria even after 40 hours of training. \tabddpm faces different problems and suffers from mode collapse, \ie it is unable to produce samples from the minority class. Training on such synthetic samples (\ie only from the majority class) yields an unusable classifier that labels all inputs as the majority class.

\para{\bgp} The \bgp task suffers from limited or biased data challenges (a)-(c) (Section~\ref{sec:data-challenges-limited}), with the notable challenge being the extremely limited training data. Training data contains 180 samples across 2 classes, with only 17 samples in the malicious class. Results are in Table~\ref{tab:genai-success-results}.

\noindent
\textsl{{\para{Finding 2}\textit{Even in extremely data challenged settings, a properly engineered GenAI scheme can provide significant performance boost.}}} Sample-conditioned GenAI scheme, \ie~\nbase, again achieves the highest performance gain, with a high mean $\Delta G$ of 32.61\%. We also see that other GenAI schemes, \nrandom and \ctabgan outperforms the traditional data augmentation schemes. These results underscore the potential of GenAI schemes.

Four GenAI schemes, \tvae, \tabsyn, \great and \realtabformer produce performance degradation. These schemes are not best suited for extremely data-challenged settings. Despite using a large pre-trained LLM backend (\ie Distill GPT-2~\cite{sanh2019distilbert}), \great and \realtabformer fail to demonstrate any gains, again highlighting that LLM-based schemes require further engineering advances before they can be practically adopted. The traditional scheme, \mcccr also results in a performance degradation. We posit that, in high-class imbalance cases, the minority point spheres become extremely small, leading to points highly similar to the original data samples, thereby likely leading to overfitting. 

\noindent
\textit{Failure cases.} No scheme failed to generate data.

\para{\bodmas}The \bodmas task has limited and bias data challenges (a), (b) and (d), along with the notable challenge of concept drift (Section~\ref{sec:data-challenges-limited}).
Concept drift can be observed by the drop in macro F-score across the 12 test months, especially for months 5 and 6 (see Figure~\ref{fig:bodmas-test}). Data from month-0 is used to train a GDBT classifier and testing is done on the subsequent 12 months of malware traces. The results are shown in Table~\ref{tab:bodmas-months2356791011} and Table~\ref{tab:bodmas-noeval-months} (Appendix).

Mitigating concept drift is a challenging problem~\cite{andresini2021insomnia}. We find that practitioners can significantly improve the performance of security classifiers before deployment by integrating GenAI. For 8/12 months, we observe a positive gain in performance, when using a GenAI scheme. In fact, we achieve over 59\% mean $\Delta G$ for month 5, where the largest impact of the concept drift was observed (Figure~\ref{fig:bodmas-test}). This is a substantial result. The results are shown in Table~\ref{tab:bodmas-months2356791011}.
For the remaining 4 months, none of the data augmentation schemes demonstrate a gain in performance (see Table~\ref{tab:bodmas-noeval-months} in the Appendix). This is expected, given the challenging nature of the problem. However, achieving performance gains in 8/12 months is a notable result. For the remainder of this section, we analyze the 8 months shown in Table~\ref{tab:bodmas-months2356791011}.

\noindent
\textsl{{\para{Finding 3}\textit{Class-conditioned GenAI approaches are well suited for tasks where there is concept drift.}}} A key finding is that class-conditioned GenAI approaches are a better fit for tasks with concept drift, compared to sample-conditioned schemes. Class-conditioning can better explore novel regions of the data manifold, where distribution shift may have occurred. This is because we are not explicitly focusing on generating samples near the vicinity of existing samples. Our class-conditioned \nrandom achieves the best performance in 6 out of the 8 months (Table~\ref{tab:bodmas-months2356791011}), \ie months 2, 3, 5, 9, 10, and 11. Detailed statistical test results to arrive at this conclusion are discussed in Section~\ref{subsec:bodmas-stat-tests}. 

On the other hand, our sample-conditioned scheme \nbase produces unreliable performance estimates in 5 out of the 8 months, and trails behind other GenAI schemes in 2 out of the remaining 3 months. Other notable schemes are \tvae and \tabddpm, which perform similar to \nrandom (no statistically significant difference) in months 9 and 10, respectively. \tvae outperforms all schemes only in month 6. 
Traditional schemes (also sample-conditioned), \smote and \mcccr perform poorly for this task. For 7 out of the 8 months, these schemes produce either negative gains or unreliable performance estimates. This highlights the limitations of traditional methods to mitigate concept drift.

\noindent
\textit{Failure scenarios.} 
Two schemes \great and \realtabformer failed to converge on \bodmas. 
\great could not be trained on this task due to its feature dimensionality exceeding 1,000. After feature processing, \great needs to be trained on over 21K tokens per sample, which far exceeds the token limit 1024 of the backend LM. \realtabformer's training failed as encoded features do not fit into the memory (NVIDIA A100 GPU).

\para{\microsoft} \microsoft, a multi-class classification task (9 classes) suffers from limited or biased training data challenges (a), (b) and (d) (Section~\ref{sec:data-challenges-limited}), with the notable challenge being the high dimensionality of features (1,804 features). Results are in Table~\ref{tab:genai-success-results}.

Class-conditioned GenAI schemes, including our new \nrandom scheme, show significant performance improvement. Amongst the class-conditioned GenAI schemes, large standard deviation values make it difficult to directly compare means. We thus perform the Kruskal-Wallis test, and find no difference in rankings of these schemes due to a p-value of 0.286 $>$ 0.05. Interestingly, our sample-conditioned scheme, \nbase, despite achieving a performance improvement, does not perform as well as the class-conditioned schemes. Similar to the \bodmas task, we suspect that this task has test samples that have drifted from the training data distribution and requires data points in more novel/diverse regions of the data manifold to achieve better performance. Recall that class-conditioning is well suited for this goal, as confirmed by the poor performance of traditional sample-conditioned schemes such as \smote and \mcccr.

\noindent
\textit{Failure scenarios.} The high dimensionality of the \microsoft task resulted in failures for the \great, \realtabformer and \tabsyn schemes.
\great suffers from issues similar to those in the \iot task, with the high-dimensionality features further exacerbating the problem. GreaT needs 17k tokens per sample for training, exceeding the backend LLM's 1024-token limit. \realtabformer and \tabsyn failed to train on the \microsoft task due to large feature encoding which does not fit into the memory (NVIDIA A100 GPU).

\para{Why does augmentation improve performance?} We perform a deep dive into \iot and \bgp tasks to understand why synthetic data improved performance. Our hypothesis is that synthetic data helped mitigate certain biases in the real dataset, thereby enhancing generalization performance. We randomly selected 50 test instances per task that \nbase's synthetic samples helped classify correctly after being misclassified before augmentation. Using Local Interpretable Model-agnostic Explanations (LIME)~\cite{ribeiro2016should}, we extract the top 10 most influential features for each corrected prediction and observe substantial distributional changes. We measure the distribution changes using entropy \ie the randomness of the probability distribution; and skewness~\cite{doane2011measuring}, a measure of the degree of asymmetry of a probability distribution. Skewness ranges from $-\infty$ to  $+\infty$, indicating the dominant tail of a distribution: negative for lower values, positive for higher, and zero for symmetry. Classifiers generalize better on symmetric distributions. 
For the \iot task, these features show a 36.8\% increase in entropy and a 93.87\% absolute reduction in skewness towards 0; for the \bgp task, entropy increases by 175.33\% and absolute reduction of skewness towards 0 by 72.8\%. For both tasks, there is a consistent reduction in the absolute value of skewness for most features after augmentation, depicted in Figure~\ref{fig:skewness-iot-bgp} (Appendix).
\textit{These shifts suggest that synthetic augmentation reduces bias and promotes more symmetric feature distributions, leading to better classifier generalization.}

\subsection{Cases where most GenAI schemes show insufficiency}
\label{subsec:genai-inadequate}

\noindent
We study 3 tasks, \tor, \cookie, and \nprintml, where we discover significant challenges to improve performance using GenAI schemes.

\para{\tor}\tor is a binary classification task that suffers from limited or biased training data challenges (a)-(c) (Section~\ref{sec:data-challenges-limited}). Table~\ref{tab:genai-noshow-results} reports the results.
\noindent
\textsl{{\para{Finding 4}\textit{Tasks with significant class overlap are challenging for GenAI-based data augmentation.}}}
No scheme except \realtabformer produces a positive gain (considering only reliable measurements). To understand this result, we visualize the t-SNE~\cite{van2008visualizing} feature representations of \tor samples and contrast them with the \bgp samples in Figure~\ref{fig:tor-tsne}. Recall that the GenAI schemes showed significant performance gains for the \bgp task. We see that \tor classes appear largely inseparable even in their non-linearly projected feature spaces, \ie with a large volume of class overlap, whereas \bgp classes have distinct separability. This indicates that GenAI data augmentation faces challenges in datasets with class overlap, often caused by suboptimal features or noisy labels.
\realtabformer is the only method to improve on the \tor task, likely due to its statistical early-stopping criterion that prevents over- and underfitting. Unlike prior settings where it failed, the lower feature dimensionality (175) and ample training samples here enabled convergence.
The success of \realtabformer highlights the promise of correctly instantiated LLM-based GenAI schemes for tasks with class overlap.

\begin{figure} [t!]
\centering
\includegraphics[height=0.18\textwidth, width=0.40\textwidth]{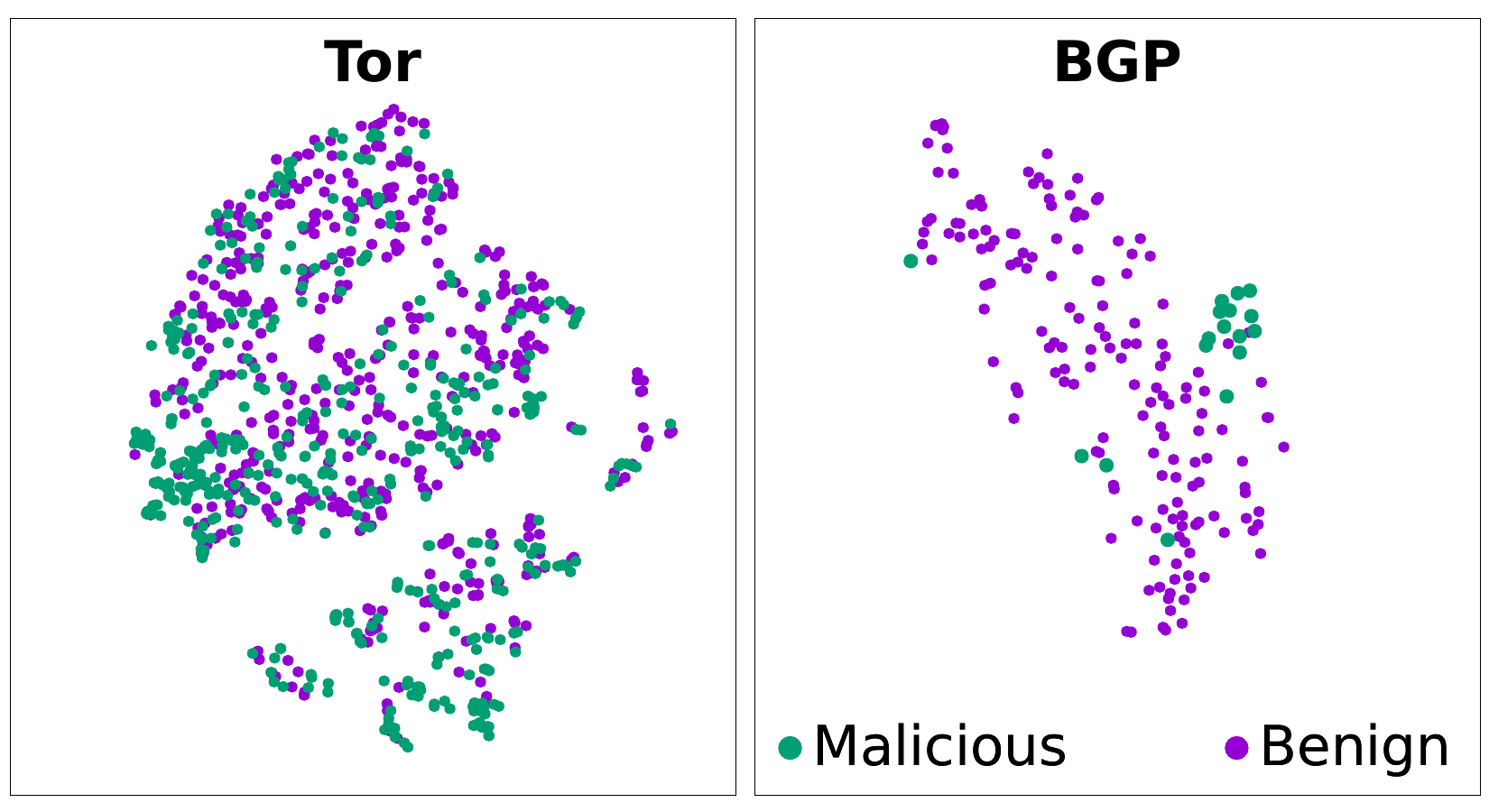}
\caption{t-SNE plot for \bgp and \tor task data. }
\Description{A t-SNE plot visualizing the data distributions for BGP and Tor tasks. The plot shows clusters representing different classes or domains, helping to illustrate the separability or overlap between them.}

\label{fig:tor-tsne}
\end{figure}
\noindent
\textit{Failure cases.} Only \great failed to converge on the \tor task. \great requires a large compute time to train on the \tor task---an estimated 22 hours per epoch (Appendix~\ref{tab:genai-train-times} (Appendix)).

\para{\cookie and \nprintml} \cookie and \nprintml are multi-class tasks with 4 and 13 classes, respectively. \cookie faces challenges (a) and (d), while \nprintml faces (a), (b), and (d) (Section~\ref{sec:data-challenges-limited}). Results are in Table~\ref{tab:genai-noshow-results}.

\noindent
\textsl{{\para{Finding 5}\textit{Tasks with noisy class labels and highly sparse feature vectors are challenging settings for GenAI schemes.}}}
Similar to the \tor task, none of the GenAI schemes demonstrate positive $\Delta G$ values on both these tasks (for reliable estimates). The only exception is the traditional scheme, \smote which shows achieves a small mean $\Delta G$ of 2.15\%.
The poor performance of GenAI schemes can be attributed to one or more of the following reasons:

\textit{1) Overlapping class distributions.} Similar to \tor, we suspect the issue of overlapping class distributions. In the \cookie task, 3,707 samples in the training set have identical feature vectors assigned to more than one class label. The authors of this work acknowledge the issue of noisy labels because ground truth labels were obtained from multiple sources that may not agree uniformly on the labels. This is a clear problem. On the other hand, for \nprintml, while we are unable to find any evidence for noisy labels, the authors hint at poor distinguishability of samples among certain classes---suggesting potential class overlap. Figure 6(b) in the original work~\cite{holland2021new}, shows poor fingerprinting performance among samples belonging different variants of the same major OS, namely Windows, and Linux. The authors attribute this problem to the different variants (of the same OS) sharing similar network stack, leading to more similarity in the extracted feature vectors.

\begin{table}[t!]
\centering
\small
\setlength{\tabcolsep}{2pt}
\setlength\extrarowheight{1.5pt}
\begin{tabular}{c"cccccc}
                                     & \multicolumn{6}{c}{\textbf{Mean $\Delta G$ with standard deviation}}                                                                                                                                                                                                                                \\ \cline{2-7} 
\multirow{-2}{*}{\textbf{Technique}} & \multicolumn{2}{c|}{\textbf{\tor}}                                                    & \multicolumn{2}{c|}{\textbf{\nprintml}}                                                       & \multicolumn{2}{c}{\textbf{\cookie}}                                     \\ \Xhline{1.1pt}
\textbf{\nbase}                      & \multicolumn{1}{c|}{-3.18 (0.00)}   &  \multicolumn{1}{c|}{\cellcolor[HTML]{da5552}$\downarrow$}  & \multicolumn{1}{c|}{0.03 (0.66)} & \multicolumn{1}{c|}{\cellcolor[HTML]{CCCCCC}-}            & \multicolumn{1}{c|}{-0.06 (0.17)} & \cellcolor[HTML]{da5552}$\downarrow$ \\ \hline
\textbf{\nrandom}                    & \multicolumn{1}{c|}{-3.87 (0.00)}   &  \multicolumn{1}{c|}{\cellcolor[HTML]{da5552}$\downarrow$} & \multicolumn{1}{c|}{-0.23 (0.57)} & \multicolumn{1}{c|}{\cellcolor[HTML]{CCCCCC}-}            & \multicolumn{1}{c|}{-0.05 (0.11)}  & \cellcolor[HTML]{da5552}$\downarrow$ \\ \hline
\textbf{\tvae}                       & \multicolumn{1}{c|}{-16.03 (6.62)}  &  \multicolumn{1}{c|}{\cellcolor[HTML]{da5552}$\downarrow$} & \multicolumn{1}{c|}{-0.68 (0.47)} & \multicolumn{1}{c|}{\cellcolor[HTML]{da5552}$\downarrow$} & \multicolumn{1}{c|}{$\times$}     & \cellcolor[HTML]{CCCCCC}$\times$     \\ \hline
\textbf{\tabsyn}                     & \multicolumn{1}{c|}{-19.54 (7.96)}  &  \multicolumn{1}{c|}{\cellcolor[HTML]{da5552}$\downarrow$}   & \multicolumn{1}{c|}{$\times$}     & \multicolumn{1}{c|}{\cellcolor[HTML]{CCCCCC}$\times$}     & \multicolumn{1}{c|}{0.11 (0.18)}  & \cellcolor[HTML]{CCCCCC}-            \\ \hline
\textbf{\great}                      & \multicolumn{1}{c|}{$\times$}     & \multicolumn{1}{c|}{\cellcolor[HTML]{CCCCCC}$\times$}   & \multicolumn{1}{c|}{$\times$}     & \multicolumn{1}{c|}{\cellcolor[HTML]{CCCCCC}$\times$}     & \multicolumn{1}{c|}{$\times$}     & \cellcolor[HTML]{CCCCCC}$\times$     \\ \hline
\textbf{\realtabformer}             & \multicolumn{1}{c|}{16.40 (6.01)}   & \multicolumn{1}{c|}{\cellcolor[HTML]{55a630}$\uparrow$} & \multicolumn{1}{c|}{$\times$}  & \multicolumn{1}{c|}{\cellcolor[HTML]{CCCCCC}$\times$} & \multicolumn{1}{c|}{$\times$}  & {\cellcolor[HTML]{CCCCCC}$\times$ } \\ \hline
\textbf{\ctabgan}                    & \multicolumn{1}{c|}{5.89 (12.52)}   &  \multicolumn{1}{c|}{\cellcolor[HTML]{CCCCCC}-}  & \multicolumn{1}{c|}{0.3 (0.50)}   & \multicolumn{1}{c|}{\cellcolor[HTML]{CCCCCC}-}            & \multicolumn{1}{c|}{0.02 (0.14)} & \cellcolor[HTML]{CCCCCC}- \\ \hline
\textbf{\tabddpm}                    & \multicolumn{1}{c|}{-21.75 (14.03)} &  \multicolumn{1}{c|}{\cellcolor[HTML]{da5552}$\downarrow$} & \multicolumn{1}{c|}{-0.22 (0.62)} & \multicolumn{1}{c|}{\cellcolor[HTML]{CCCCCC}-}            & \multicolumn{1}{c|}{-0.01 (0.11)} & \cellcolor[HTML]{da5552}$\downarrow$ \\ \hline
\textbf{\smote}                      & \multicolumn{1}{c|}{4.3 (5.12)}     &  \multicolumn{1}{c|}{\cellcolor[HTML]{CCCCCC}-}  & \multicolumn{1}{c|}{-0.3 (0.58)}  & \multicolumn{1}{c|}{\cellcolor[HTML]{CCCCCC}-}            & \multicolumn{1}{c|}{2.15 (0.08)}  & \cellcolor[HTML]{55a630}$\uparrow$   \\ \hline
\textbf{\mcccr}                      & \multicolumn{1}{c|}{-30.06 (0.00)}  &  \multicolumn{1}{c|}{\cellcolor[HTML]{da5552}$\downarrow$}         & \multicolumn{1}{c|}{0.37 (0.55)}  & \multicolumn{1}{c|}{\cellcolor[HTML]{CCCCCC}-}            & \multicolumn{1}{c|}{-0.04 (0.10)} & \cellcolor[HTML]{da5552}$\downarrow$ \\ 
\end{tabular}
\caption{Mean $\Delta G$ with standard deviation for \tor, \nprintml and \cookie tasks, where GenAI schemes do not show performance improvement. $\uparrow$ and $\downarrow$ indicate positive and negative gains, respectively. `-' indicates unreliable gains (CV $\geq$ 1), and `$\times$' indicates model failures.}
\label{tab:genai-noshow-results}
\end{table}

\textit{2) Sparse feature vectors.} The training datasets of \nprintml and \cookie show high feature sparsity (percentage of feature values that are zero)—78.73\% and 98.66\%, respectively. To assess the impact of sparsity on performance, we run a new experiment, minimizing potential class overlap by reducing the 13-class \nprintml task to 5 carefully selected classes.\footnote{We only retain 5 OS classes - i.e. Mac, Kali-linux, Ubuntu-desktop (combining all variants), Ubuntu-server and Windows (combining all variants).}
\smote applied to this setting produced a mean $\Delta G$ of -0.31\%, showing no improvement, suggesting sparsity plays a role in the non-improvement of performance. To further examine whether GenAI schemes struggle with performance improvement in noisy/overlapping data, we artificially induce class overlap in \bgp. Recall that GenAI schemes (\nbase and \ctabgan) performed well on \bgp. We flip 20\% of class 0 samples to class 1 to induce overlap. This resulted in these GenAI methods yielding negative gains of up to -37\%. This further strengthens our \textit{Finding 5}. More details are in Appendix~\ref{subsec:noisy-bgp-effect}.

\noindent
\textit{Failure scenarios.} Four GenAI schemes, \great, \realtabformer, \tvae and \tabsyn failed in one or both tasks. We were unable to train \great for both tasks. Both tasks have feature dimensionality over 1,000, resulting in over 11K and 28K tokens per sample, which is inordinately greater than the token limit of 1024 of the backend LLMs.
\realtabformer failed to train on both tasks. For \nprintml, training fails due to large encoded features which does not fit into the memory (NVIDIA A100 GPU). For \cookie, training does not converge as our early stopping criteria is not met even after 40 hours.}
\tvae failed for the \cookie task, because we failed to generate new samples, even after 100 hours of rejection sampling (see Appendix~\ref{subsec:genai-additional}). \tabsyn failed for the \nprintml task. \tabsyn’s VAE fails to fit in GPU memory (NVIDIA A100) for \nprintml due to its large feature size and many classes.

\section{Evaluation: Post-Deployment}
\label{sub-sec:concept-drift-evaluation}
\noindent GenAI schemes are still limited by existing training data in improving classifier generalization. We again focus on a concept drift setting, where GenAI schemes may struggle to generate points that fit the drifted test. To mitigate drift after deployment, the traditional practice involves a costly and time-consuming labeling process for the drifted test set, followed by retraining the classifier on the new data. Security threats are known to change abruptly~\cite{andresini2021insomnia}, requiring a faster reaction (\eg for NIDS and malware detection). In this section, we investigate whether GenAI can be used to perform \textit{faster recovery} from concept drift, without requiring a complete data labeling effort on the test set. To achieve this, we propose a novel scheme called \textit{\sysname-hybrid} that combines both the \nrandom and \nbase approaches.

Similar to the setup in Section~\ref{subsubsec:genai-potential}, we train our \sysname scheme on month 0 data. We focus to mitigate concept drift on months 5 and 6, which show the highest performance drop (Figure~\ref{fig:bodmas-test}). We assume the defender has detected the concept drift in month $k$ ($k \in [1,5]$). Existing methods such as CADE~\cite{yang2021cade} and Transcend~\cite{jordaney2017transcend, barbero2022transcending} can be used to detect concept drift. To rapidly recover from concept drift, we use \nbase's sample conditioning to generate synthetic points from a small labeled subset of the drifted test set from month $k$. This is possible as \sysname can perform sample-conditioned generation on new samples without the need for retraining, unlike other GenAI techniques that perform class-conditioned generation. The samples for labeling are chosen using uncertanity sampling~\cite{andresini2021insomnia}, an active learning technique~\cite{aggarwal2014active}. It chooses samples close to the decision boundary based on classifier confidence scores. We show that even a small subset is sufficient to boost performance via data augmentation, with subset sizes ranging from 9 to 64 for the \BODMAS task.
This significantly reduces the cost of labeling effort, speeding adaptation to drift, \ie fast recovery.

For a testing month $N$ ($N$ can be months 5 and 6), we evaluate whether uncertainty samples from any of the previous $(N-1)$ months can improve performance.
We use a confidence interval of 0.3-0.7 from a random 1\% subset of the test set to determine uncertainty samples. We augment \nbase with 5 times the synthetic samples conditioned on uncertain samples. The remaining synthetic samples are generated using \nrandom (\ie class-conditioned) to balance the class sizes.

We compare \sysname-hybrid with the three baseline schemes: (1) \textit{w/o GenAI}: This reports classifier performance without any data augmentation, \ie only using the real data. 
(2) \textit{\nrandom}: This mirrors the setting in Section~\ref{subsubsec:genai-potential}, where class conditioning uses only month 0 data for generating synthetic samples.
(3) \textit{INSOMNIA}: The security classifier is retrained on a combination of real and uncertainty samples (the same samples used by \sysname-hybrid). This strategy is equivalent to the existing method called INSOMNIA~\cite{andresini2021insomnia} which aims to adapt classifiers to concept drift using uncertainty samples. We use the macro F-score, reporting the mean and standard deviation across 10 trials for \nrandom and \sysname-hybrid.

\begin{table}[t!]
\centering
\small
\setlength{\tabcolsep}{8pt}
\setlength\extrarowheight{1.5pt}
\begin{tabular}{l"ll}
\multirow{2}{*}{\textbf{Method}} & \multicolumn{2}{c}{\textbf{\begin{tabular}[c]{@{}c@{}}F-score for months \\ (w/ standard deviation)\end{tabular}}} \\ \cline{2-3} 
                      & \multicolumn{1}{c|}{\textbf{5}}   & \multicolumn{1}{c}{\textbf{6}} \\ \Xhline{1.1pt}
\textbf{w/o GenAI}    & \multicolumn{1}{l|}{36.40 (-)}    & 33.60 (-)                      \\ \hline
\textbf{\nrandom}     & \multicolumn{1}{l|}{58.03 (3.00)} & 43.09 (6.77)                   \\ \hline
\textbf{INSOMNIA}     & \multicolumn{1}{l|}{34.40 (-)}    & 33.40 (-)                      \\ \hline
\textbf{\sysname-hybrid} & \multicolumn{1}{l|}{59.78 (8.34)} & \textbf{53.92 (10.39)}        
\end{tabular}
\caption{Mean macro F-scores and standard deviation for \nrandom and \sysname-hybrid across 10 trials and macro F-scores for w/o GenAI and INSOMNIA on \bodmas months 5 \& 6. (-) indicates no standard deviation.}
\label{tab:concept-drift}
\end{table}

\noindent
\textsl{{\para{Finding 6}\textit{A combination of class and sample-conditioned GenAI schemes can help to quickly recover from concept drift with low-cost labeling effort.}}} For months 5 and 6, we present the best performance obtained by any of the previous $(N-1)$ months across 10 trials of data augmentation. Results for months 5 and 6 are shown in Table~\ref{tab:concept-drift}. Complete results are in Table~\ref{tab:concept-drift-all-months} (Appendix). High deviations led us to use the Kruskal–Wallis test for ranking in both months. In month 5, \sysname-hybrid and \nrandom show no difference in means, making them equivalent schemes. In month 6, which shows the maximum drop in performance in all test months of \bodmas, we find \sysname-hybrid outperforms \nrandom. This shows the potential for \sysname to rapidly recover using drifting samples post-deployment. We conducted our experiment for the remaining 10 months (which did not degrade as much as months 5 and 6) but observed no performance improvement from using \sysname-hybrid, indicating room for advancing our recovery strategy. 
However, we provided significant gains for the worst affected months.

\section{Conclusion \& Future Work}
\label{sec:conclusion-future-work}
\noindent
We now address questions in Section~\ref{sec:goals} using our findings in Section~\ref{sub-sec:genai-eval} and also suggest future directions.

\para{Challenges with instantiating GenAI models} Unlike traditional data augmentation schemes (\eg SMOTE), GenAI methods are highly complex ML schemes, which are challenging to correctly instantiate for wide diversity of security tasks. 5/6 GenAI schemes fail to generate data for at least 1/7 security tasks. \tabddpm suffered from mode collapse, \tvae failed to generate sufficient samples while class-conditioning, and \tabsyn is unable to manage large, high-dimensional datasets. While \realtabformer shows potential in \tor task, both \great and \realtabformer fail on most tasks (6/7 and 5/7, respectively) due to limitations with high-dimensional feature vectors. Re-engineering LLM-based approaches like \great and \realtabformer for greater scalability with large, high-dimensional security datasets offers several benefits: (1) LLMs have robust methods for controlled text generation which could help to design effective sample-conditioning schemes, (2) potential for prompt-based data generation via instruction tuning~\cite{Chung2022ScalingIL}, (3) ability to handle complex data types (discrete or continuous) as everything is represented in text, and (4) ability to augment in overlapping distributions.

\para{GenAI-based data augmentation to boost performance}Data augmentation using GenAI is a promising approach to boost performance of diverse security classifiers---we improved performance for 5/7 security tasks, without any changes to the underlying classification algorithm. Therefore, a carefully engineered GenAI tool can reignite progress for security problems stagnated by slow algorithmic innovations---\ie by simply addressing any underlying data challenges.  Future work can explore more challenging scenarios, including overlapping class distributions, noisy labels, sparse feature vectors. Developments in LLMs~\cite{wang2024llmautoda, li-etal-2024-search}, Diffusion models~\cite{samuel2024generating, park2025raretofrequent, shao2024diffult}, VAE~\cite{stocksieker2024data, fajardo2018vos, kim2024tvariational} and GANs~\cite{khorram2024taming} such as strategic sampling methods and weighted learning approaches
can be leveraged to produce rare/limited data points from training data manifolds.

\para{Impact of highly controlled generation} Existing GenAI schemes do not provide mechanisms for highly controlled generation. To fill this gap, we proposed \nbase, a novel scheme based on a VAE using a discrete latent space to enable generation of samples targeting specific regions of the data distribution. This sample-conditioned approach achieved the best performance in 3 security tasks with complex biases. Future work can explore approaches like LLMs, Diffusion models, and GANs.

\para{Rapid recovery from concept drift}We showed that GenAI can help maintain classifier performance post-deployment through a malware classification task exhibiting concept drift. A hybrid \sysname scheme, combining sample- and class-conditioning, enabled rapid recovery with a few labeled samples from the drifted set. While effective, our recovery approach has room for improvement. Future work could explore using GenAI to recover from adversarial attacks. 

We discuss the ethical considerations of this work in Section~\ref{sec:ethics} (Appendix).

\begin{acks} 
\noindent This material is based upon work supported by the National Science Foundation under grant award numbers 2453818, 2453819 and 2453820. Neal Mangaokar was supported by an NSF Graduate Fellowship.
\end{acks}

\bibliographystyle{ACM-Reference-Format}
\bibliography{main.bib}

@STRING{cvpr	= "Proc. of {CVPR}" }

@STRING{ndss	= "Proc. of {NDSS}" }

@STRING{springer= "Springer-Verlag" }

@inproceedings{yin2022practical,
  title={Practical gan-based synthetic ip header trace generation using netshare},
  author={Yin, Yucheng and Lin, Zinan and Jin, Minhao and Fanti, Giulia and Sekar, Vyas},
  booktitle={Proc. of SIGCOMM},
  year={2022}
}

@article{jiang2024netdiffusion,
  title={Netdiffusion: Network data augmentation through protocol-constrained traffic generation},
  author={Jiang, Xi and others},
  journal={POMACS},
  year={2024}
}

@inproceedings{van2016conditional,
  title={{Conditional Image Generation with PixelCNN Decoders}},
  author={Van den Oord, Aaron and Kalchbrenner, Nal and Espeholt, Lasse and Vinyals, Oriol and Graves, Alex and others},
  booktitle={Proc. of NeurIPS}, 
  year={2016}
}

@book{bishop2006pattern,
  title={Pattern recognition and machine learning},
  author={Bishop, Christopher M and Nasrabadi, Nasser M},
  year={2006},
  publisher={Springer}
}

@inproceedings{taylor2016anomaly,
  title={{Anomaly Detection in Automobile Control Network Data with Long Short-Term Memory Networks}},
  author={Taylor, Adrian and Leblanc, Sylvain and Japkowicz, Nathalie},
  booktitle={Proc. of IEEE DSAA}, 
  year={2016} 
}

@inproceedings{goodfellow2014generative,
  title={{Generative Adversarial Nets}},
  author={Goodfellow, Ian and Pouget-Abadie, Jean and Mirza, Mehdi and Xu, Bing and Warde-Farley, David and Ozair, Sherjil and Courville, Aaron and Bengio, Yoshua},
  booktitle={Proc. of NeurIPS}, 
  year={2014}
}

@article{mc-ccr,
  title={{Combined Cleaning and Resampling Algorithm for Multi-Class Imbalanced Data with Label Noise}},
  author={Koziarski, Micha{\l} and Wo{\'z}niak, Micha{\l} and Krawczyk, Bartosz},
  journal={Knowledge-Based Systems}, 
  year={2020} 
}

@inproceedings{dathathri2019plug,
  title={{Plug and Play Language Models: A Simple Approach to Controlled Text Generation}},
  author={Dathathri, Sumanth and Madotto, Andrea and Lan, Janice and Hung, Jane and Frank, Eric and Molino, Piero and Yosinski, Jason and Liu, Rosanne},
  booktitle={Proc. of ICLR},
  year={2020}
}

@inproceedings{hayes2016k,
  title={{k-fingerprinting: a Robust Scalable Website Fingerprinting Technique}},
  author={Hayes, Jamie and Danezis, George},
  booktitle={Proc. of USENIX Security}, 
  year={2016}
}

@article{krawczyk2019radial,
  title={{Radial-Based Oversampling for Multiclass Imbalanced Data Classification}},
  author={Krawczyk, Bartosz and Koziarski, Micha{\l} and Wo{\'z}niak, Micha{\l}},
  journal={IEEE TNNLS}, 
  year={2019} 
}

@inproceedings{cicids2017,
  title={{Toward Generating a New Intrusion Detection Dataset and Intrusion Traffic Characterization}},
  author={Sharafaldin, Iman and Lashkari, Arash Habibi and Ghorbani, Ali A and others},
  booktitle={Proc. of ICISSP}, 
  year={2018}
}

@article{lu2022generative,
  title={{Generative Adversarial Networks (GANs) for Image Augmentation in Agriculture: A Systematic Review}},
  author={Lu, Yuzhen and Chen, Dong and Olaniyi, Ebenezer and Huang, Yanbo},
  journal={Comput. Electron. Agric.}, 
  year={2022} 
}

@article{chlap2021review,
  title={{A Review of Medical Image Data Augmentation Techniques for Deep Learning Applications}},
  author={Chlap, Phillip and others},
  journal={JMIRO}, 
  year={2021} 
}

@inproceedings{dodia2022exposing,
  title={{Exposing the Rat in the Tunnel: Using Traffic Analysis for Tor-based Malware Detection}},
  author={Dodia, Priyanka and AlSabah, Mashael and Alrawi, Omar and Wang, Tao},
  booktitle={Proc. of CCS}, 
  year={2022}
}

@InProceedings{ribeiro2016should,
  title		= {Why should i trust you?: Explaining the predictions of any
		  classifier},
  author	= {Ribeiro, Marco Tulio and Singh, Sameer and Guestrin,
		  Carlos},
  booktitle	= {Proc. of KDD},
  year		= {2016}
}

@inproceedings{samuel2024generating,
  title={Generating images of rare concepts using pre-trained diffusion models},
  author={Samuel, Dvir and Ben-Ari, Rami and Raviv, Simon and Darshan, Nir and Chechik, Gal},
  booktitle={Proc. of AAAI},
  year={2024}
}

@article{fajardo2018vos,
  title={Vos: a method for variational oversampling of imbalanced data},
  author={Fajardo, Val Andrei and Findlay, David and Houmanfar, Roshanak and Jaiswal, Charu and Liang, Jiaxi and Xie, Honglei},
  journal={Proc. of CoRR abs/1809.02596},
  year={2018}
}

@inproceedings{
kim2024tvariational,
title={\$t{\textasciicircum}3\$-Variational Autoencoder: Learning Heavy-tailed Data with Student's t and Power Divergence},
author={Juno Kim and Jaehyuk Kwon and Mincheol Cho and Hyunjong Lee and Joong-Ho Won},
booktitle={Proc. of ICLR},
year={2024}
}

@article{stocksieker2024data,
  title={Data Augmentation with Variational Autoencoder for Imbalanced Dataset},
  author={Stocksieker, Samuel and Pommeret, Denys and Charpentier, Arthur},
  journal={Proc. of CoRR abs/2412.07039},
  year={2024}
}

@inproceedings{nong2024vgx,
  title={VGX: Large-Scale Sample Generation for Boosting Learning-Based Software Vulnerability Analyses},
  author={Nong, Yu and others},
  booktitle={Proc. of ICSE},
  year={2024}
}

@inproceedings{qiu2021deepsweep,
  title={Deepsweep: An evaluation framework for mitigating DNN backdoor attacks using data augmentation},
  author={Qiu, Han and Zeng, Yi and Guo, Shangwei and Zhang, Tianwei and Qiu, Meikang and Thuraisingham, Bhavani},
  booktitle={Proc. of Asia CCS},
  year={2021}
}

@article{Chung2022ScalingIL,
  title={{Scaling Instruction-Finetuned Language Models}},
  author={Hyung Won Chung and others},
  year={2022},
  journal={Proc. of CoRR abs/2210.11416},
}

@inproceedings{bai2024sequential,
  title={Sequential modeling enables scalable learning for large vision models},
  author={Bai, Yutong and others},
  booktitle={Proc. of CVPR},
  year={2024}
}

@inproceedings{sanh2019distilbert,
  title={{DistilBERT, a distilled version of BERT: smaller, faster, cheaper and lighter}},
  author={Sanh, Victor and Debut, Lysandre and Chaumond, Julien and Wolf, Thomas},
  booktitle={Proc. of NeurIPS},
  year={2019}
}

@article{abdi2010coefficient,
  title={Coefficient of variation},
  author={Abdi, Herv{\'e}},
  journal={Encyclopedia of research design},
  year={2010}
}

@article{dunn1964multiple,
  title={{Multiple Comparisons Using Rank Sums}},
  author={Dunn, Olive Jean},
  journal={Technometrics}, 
  year={1964} 
}

@article{van2008visualizing,
  title={{Visualizing Data using t-SNE}},
  author={Van der Maaten, Laurens and Hinton, Geoffrey},
  journal={JMLR}, 
  year={2008}
}

@book{montgomery2019design,
  title={Design and Analysis of Experiments},
  author={Montgomery, Douglas C.},
  year={2019},
  publisher={John Wiley \& Sons},
  edition={10th},
}

@article{kruskal1952use,
  title={{Use of Ranks in One-Criterion Variance Analysis}},
  author={Kruskal, William H and Wallis, W Allen},
  journal={JASA}, 
  year={1952} 
}

@inproceedings{ahmadi2016novel, 
title={{Novel Feature Extraction, Selection and Fusion for Effective Malware Family Classification}}, 
author={Ahmadi, Mansour and Ulyanov, Dmitry and Semenov, Stanislav and Trofimov, Mikhail and Giacinto, Giorgio}, 
booktitle={Proc. Of ACM CODASPY}, 
year={2016} 
}

@inproceedings{andresini2021insomnia,
  title={{INSOMNIA: Towards Concept-Drift Robustness in Network Intrusion Detection}},
  author={Andresini, Giuseppina and Pendlebury, Feargus and Pierazzi, Fabio and Loglisci, Corrado and Appice, Annalisa and Cavallaro, Lorenzo},
  booktitle={Proc. of ACM AISec}, 
  year={2021}
}

@inproceedings{lin2020using,
  title={{Using GANs for Sharing Networked Time Series Data: Challenges, Initial Promise, and Open Questions}},
  author={Lin, Zinan and Jain, Alankar and Wang, Chen and Fanti, Giulia and Sekar, Vyas},
  booktitle={Proc. of ACM IMC}, 
  year={2020}
}

@article{keskar2019ctrl,
  title={{CTRL: A Conditional Transformer Language Model for Controllable Generation}},
  author={Keskar, Nitish Shirish and McCann, Bryan and Varshney, Lav R and Xiong, Caiming and Socher, Richard},
  journal={CoRR abs/1909.05858},
  year={2019}
}

@article{christ2018time,
  title={{Time Series FeatuRe Extraction on basis of Scalable Hypothesis tests (tsfresh – A Python package)}},
  author={Christ, Maximilian and Braun, Nils and Neuffer, Julius and Kempa-Liehr, Andreas W},
  journal={Neurocomputing}, 
  year={2018} 
}

@article{he2021automl,
  title={{AutoML: A Survey of the State-of-the-Art}},
  author={He, Xin and Zhao, Kaiyong and Chu, Xiaowen},
  journal={Knowledge-Based Systems}, 
  year={2021} 
}

@MISC {LIEF,
  author       = "Romain Thomas",
  title        = "LIEF - Library to Instrument Executable Formats",
  howpublished = "https://lief.quarkslab.com/",
  year         = "2017"
}

@misc{kaggle_malware_classification,
author = {},
title = {{Microsoft Malware Classification Challenge - Kaggle}},
howpublished = {\url{https://www.kaggle.com/c/malware-classification}},
month = {},
year = {2015} 
}

@article{smote,
  title={{SMOTE: Synthetic Minority Over-sampling Technique}},
  author={Chawla, Nitesh V and Bowyer, Kevin W and Hall, Lawrence O and Kegelmeyer, W Philip},
  journal={JAIR}, 
  year={2002}
}

@inproceedings{adasyn,
  title={{ADASYN: Adaptive Synthetic Sampling Approach for Imbalanced Learning}},
  author={He, Haibo and Bai, Yang and Garcia, Edwardo A and Li, Shutao},
  booktitle={Proc. of IEEE WCCI}, 
  year={2008} 
}

@article{deerwester1990indexing,
  title={{Indexing by Latent Semantic Analysis}},
  author={Deerwester, Scott and Dumais, Susan T and Furnas, George W and Landauer, Thomas K and Harshman, Richard},
  journal={JASIST}, 
  year={1990} 
}

@article{doane2011measuring,
  title={Measuring skewness: a forgotten statistic?},
  author={Doane, David P and Seward, Lori E},
  journal={Journal of statistics education},
  year={2011},
}

@inproceedings{xu2021deep,
  title={{Deep Entity Classification: Abusive Account Detection for Online Social Networks}},
  author={Xu, Teng and others},
  booktitle={Proc. of USENIX Security},
  year={2021}
}

@inproceedings{chen2020training,
  title={{On Training Robust PDF Malware Classifiers}},
  author={Chen, Yizheng and Wang, Shiqi and She, Dongdong and Jana, Suman},
  booktitle={Proc. of USENIX Security}, 
  year={2020}
}

@inproceedings{sikder20176thsense,
  title={{6thSense: A Context-aware Sensor-based Attack Detector for Smart Devices}},
  author={Sikder, Amit Kumar and Aksu, Hidayet and Uluagac, A Selcuk},
  booktitle={Proc. of USENIX Security}, 
  year={2017}
}

@inproceedings{cidon2019high,
  title={{High Precision Detection of Business Email Compromise}},
  author={Cidon, Asaf and Gavish, Lior and Bleier, Itay and Korshun, Nadia and Schweighauser, Marco and Tsitkin, Alexey},
  booktitle={Proc. of USENIX Security}, 
  year={2019}
}

@inproceedings{yu2020you,
  title={{You Are What You Broadcast: Identification of Mobile and IoT Devices from (Public) WiFi}},
  author={Yu, Lingjing and Luo, Bo and Ma, Jun and Zhou, Zhaoyu and Liu, Qingyun},
  booktitle={Proc. of USENIX Security}, 
  year={2020}
}

@inproceedings{ho2019detecting,
  title={{Detecting and Characterizing Lateral Phishing at Scale}},
  author={Ho, Grant and Cidon, Asaf and Gavish, Lior and Schweighauser, Marco and Paxson, Vern and Savage, Stefan and Voelker, Geoffrey M and Wagner, David},
  booktitle={Proc. of USENIX Security}, 
  year={2019}
}

@inproceedings{schuppen2018fanci,
  title={{FANCI : Feature-based Automated NXDomain Classification and Intelligence}},
  author={Sch{\"u}ppen, Samuel and Teubert, Dominik and Herrmann, Patrick and Meyer, Ulrike},
  booktitle={Proc. of USENIX Security}, 
  year={2018}
}

@inproceedings{barradas2018effective,
  title={{Effective Detection of Multimedia Protocol Tunneling using Machine Learning}},
  author={Barradas, Diogo and Santos, Nuno and Rodrigues, Lu{\'\i}s},
  booktitle={Proc. of USENIX Security}, 
  year={2018}
}

@inproceedings{banescu2017predicting,
  title={{Predicting the Resilience of Obfuscated Code Against Symbolic Execution Attacks via Machine Learning}},
  author={Sebastian, B and Christian, C and Alexander, P},
  booktitle={Proc. of USENIX Security}, 
  year={2017}
}

@inproceedings{jordaney2017transcend,
  title={{Transcend: Detecting Concept Drift in Malware Classification Models}},
  author={Jordaney, Roberto and Sharad, Kumar and Dash, Santanu K and Wang, Zhi and Papini, Davide and Nouretdinov, Ilia and Cavallaro, Lorenzo},
  booktitle={Proc. of USENIX Security}, 
  year={2017}
}

@inproceedings{downing2021deepreflect,
  title={{DeepReflect: Discovering Malicious Functionality through Binary Reconstruction}},
  author={Downing, Evan and Mirsky, Yisroel and Park, Kyuhong and Lee, Wenke},
  booktitle={Proc. of USENIX Security}, 
  year={2021}
}

@inproceedings{guo2019deepvsa,
  title={{DEEPVSA: Facilitating Value-set Analysis with Deep Learning for Postmortem Program Analysis}},
  author={Guo, Wenbo and Mu, Dongliang and Xing, Xinyu and Du, Min and Song, Dawn},
  booktitle={Proc. of USENIX Security}, 
  year={2019}
}

@inproceedings{han2021sigl,
  title={{SIGL: Securing Software Installations Through Deep Graph Learning}},
  author={Han, Xueyuan and others},
  booktitle={Proc. of USENIX Security}, 
  year={2021}
}

@inproceedings{xu2019anatomy,
  title={{The Anatomy of a Cryptocurrency Pump-and-Dump Scheme}},
  author={Xu, Jiahua and Livshits, Benjamin},
  booktitle={Proc. of USENIX Security}, 
  year={2019}
}

@inproceedings{yang2021cade,
  title={{CADE: Detecting and Explaining Concept Drift Samples for Security Applications}},
  author={Yang, Limin and others},
  booktitle={Proc. of USENIX Security}, 
  year={2021}
}

@inproceedings{chen2021cost,
  title={{Cost-Aware Robust Tree Ensembles for Security Applications}},
  author={Chen, Yizheng and Wang, Shiqi and Jiang, Weifan and Cidon, Asaf and Jana, Suman},
  booktitle={Proc. of USENIX Security}, 
  year={2021}
}

@inproceedings{yang2022wtagraph,
  title={{WTAGRAPH: Web Tracking and Advertising Detection using Graph Neural Networks}},
  author={Yang, Zhiju and Pei, Weiping and Chen, Monchu and Yue, Chuan},
  booktitle={Proc. of IEEE S\&P}, 
  year={2022} 
}

@inproceedings{jan2020throwing,
  title={{Throwing Darts in the Dark? Detecting Bots with Limited Data using Neural Data Augmentation}},
  author={Jan, Steve TK and Hao, Qingying and Hu, Tianrui and Pu, Jiameng and Oswal, Sonal and Wang, Gang and Viswanath, Bimal},
  booktitle={Proc. of IEEE S\&P}, 
  year={2020} 
}

@inproceedings{chen2018learning,
  title={{Learning from Mutants: Using Code Mutation to Learn and Monitor Invariants of a Cyber-Physical System}},
  author={Chen, Yuqi and Poskitt, Christopher M and Sun, Jun},
  booktitle={Proc. of IEEE S\&P}, 
  year={2018} 
}

@inproceedings{barbero2022transcending,
  title={{Transcending TRANSCEND: Revisiting Malware Classification in the Presence of Concept Drift}},
  author={Barbero, Federico and Pendlebury, Feargus and Pierazzi, Fabio and Cavallaro, Lorenzo},
  booktitle={Proc. of IEEE S\&P}, 
  year={2022} 
}

@inproceedings{li2018machine,
  title={{A Machine Learning Approach To Prevent Malicious Calls Over Telephony Networks}},
  author={Li, Huichen and Xu, Xiaojun and Liu, Chang and Ren, Teng and Wu, Kun and Cao, Xuezhi and Zhang, Weinan and Yu, Yong and Song, Dawn},
  booktitle={Proc. of IEEE S\&P}, 
  year={2018} 
}

@inproceedings{she2020neutaint,
  title={{Neutaint: Efficient Dynamic Taint Analysis with Neural Networks}},
  author={She, Dongdong and Chen, Yizheng and Shah, Abhishek and Ray, Baishakhi and Jana, Suman},
  booktitle={Proc. of IEEE S\&P}, 
  year={2020} 
}

@inproceedings{she2019neuzz,
  title={{NEUZZ: Efficient Fuzzing with Neural Program Smoothing}},
  author={She, Dongdong and Pei, Kexin and Epstein, Dave and Yang, Junfeng and Ray, Baishakhi and Jana, Suman},
  booktitle={Proc. of IEEE S\&P}, 
  year={2019} 
}

@inproceedings{thirumuruganathan2022siraj,
  title={{SIRAJ: A Unified Framework for Aggregation of Malicious Entity Detectors}},
  author={Thirumuruganathan, Saravanan and Nabeel, Mohamed and Choo, Euijin and Khalil, Issa and Yu, Ting},
  booktitle={Proc. of IEEE S\&P}, 
  year={2022} 
}

@inproceedings{peeters2018sonar,
  title={{Sonar: Detecting SS7 Redirection Attacks With Audio-Based Distance Bounding}},
  author={Peeters, Christian and Abdullah, Hadi and Scaife, Nolen and Bowers, Jasmine and Traynor, Patrick and Reaves, Bradley and Butler, Kevin},
  booktitle={Proc. of IEEE S\&P}, 
  year={2018} 
}

@inproceedings{gong2022surakav,
  title={{Surakav: Generating Realistic Traces for a Strong Website Fingerprinting Defense}},
  author={Gong, Jiajun and Zhang, Wuqi and Zhang, Charles and Wang, Tao},
  booktitle={Proc. of IEEE S\&P}, 
  year={2022} 
}

@inproceedings{yang2021bodmas,
  title={{BODMAS: An Open Dataset for Learning based Temporal Analysis of PE Malware}},
  author={Yang, Limin and Ciptadi, Arridhana and Laziuk, Ihar and Ahmadzadeh, Ali and Wang, Gang},
  booktitle={Proc. of IEEE S\&P Workshop}, 
  year={2021} 
}

@inproceedings{bollinger2022automating,
  title={{Automating Cookie Consent and GDPR Violation Detection}},
  author={Bollinger, Dino and Kubicek, Karel and Cotrini, Carlos and Basin, David},
  booktitle={Proc. of USENIX Security}, 
  year={2022}
}

@inproceedings{zhou2019devign,
  title={{Devign: Effective Vulnerability Identification by Learning Comprehensive Program Semantics via Graph Neural Networks}},
  author={Zhou, Yaqin and Liu, Shangqing and Siow, Jingkai and Du, Xiaoning and Liu, Yang},
  booktitle={Proc. of NeurIPS}, 
  year={2019}
}

@inproceedings{park2019empirical,
  title={{An Empirical Study of Prioritizing JavaScript Engine Crashes via Machine Learning}},
  author={Park, Sunnyeo and Kim, Dohyeok and Son, Sooel},
  booktitle={Proc. of AsiaCCS}, 
  year={2019}
}

@inproceedings{tekiner2022lightweight,
  title={{A Lightweight IoT Cryptojacking Detection Mechanism in Heterogeneous Smart Home Networks}},
  author={Tekiner, Ege and Acar, Abbas and Uluagac, A Selcuk},
  booktitle={Proc. of NDSS},
  year={2022}
}

@inproceedings{holland2021new,
  title={{New Directions in Automated Traffic Analysis}},
  author={Holland, Jordan and Schmitt, Paul and Feamster, Nick and Mittal, Prateek},
  booktitle={Proc. of CCS}, 
  year={2021}
}

@article{zhu2021botspot++,
  title={{BotSpot++: A Hierarchical Deep Ensemble Model for Bots Install Fraud Detection in Mobile Advertising}},
  author={Zhu, Yadong and Wang, Xiliang and Li, Qing and Yao, Tianjun and Liang, Shangsong},
  journal={ACM TOIS}, 
  year={2021} 
}

@inproceedings{testart2019profiling,
  title={{Profiling BGP Serial Hijackers: Capturing Persistent Misbehavior in the Global Routing Table}},
  author={Testart, Cecilia and Richter, Philipp and King, Alistair and Dainotti, Alberto and Clark, David},
  booktitle={Proc. of ACM IMC}, 
  year={2019}
}

@inproceedings{mirsky5kitsune,
  title={{Kitsune: An Ensemble of Autoencoders for Online Network Intrusion Detection}},
  author={Mirsky, Yisroel and Doitshman, Tomer and Elovici, Yuval and Shabtai, Asaf},
  booktitle={Proc. of NDSS},
  year={2018}
}

@article{bengio2000taking,
  title={{Taking on the Curse of Dimensionality in Joint Distributions Using Neural Networks}},
  author={Bengio, Samy and Bengio, Yoshua},
  journal={IEEE TNN}, 
  year={2000} 
}

@article{blagus2013smote,
  title={{SMOTE for high-dimensional class-imbalanced data}},
  author={Blagus, Rok and Lusa, Lara},
  journal={BMC Bioinformatics}, 
  year={2013} 
}

@article{fernandez2018smote,
  title={{SMOTE for Learning from Imbalanced Data: Progress and Challenges, Marking the 15-year Anniversary}},
  author={Fern{\'a}ndez, Alberto and Garcia, Salvador and Herrera, Francisco and Chawla, Nitesh V},
  journal={JAIR}, 
  year={2018}
}

@article{dablain2022deepsmote,
  title={{DeepSMOTE: Fusing Deep Learning and SMOTE for Imbalanced Data}},
  author={Dablain, D and Krawczyk, B and DeepSMOTE, NV Chawla},
  journal={IEEE TNNLS}, 
  year={2021}
}

@inproceedings{sdv,
  title={{The Synthetic Data Vault}},
  author={Patki, Neha and Wedge, Roy and Veeramachaneni, Kalyan},
  booktitle={Proc. of IEEE DSAA}, 
  year={2016} 
}

@inproceedings{dai2019generative,
  title={{Generative Oversampling with a Contrastive Variational Autoencoder}},
  author={Dai, Wangzhi and Ng, Kenney and Severson, Kristen and Huang, Wei and Anderson, Fred and Stultz, Collin},
  booktitle={Proc. of ICDM}, 
  year={2019}
}

@inproceedings{vqvae,
  title={{Neural Discrete Representation Learning}},
  author={Van Den Oord, Aaron and Vinyals, Oriol and others},
  booktitle={Proc. of NeurIPS}, 
  year={2017}
}

@article{aneja2021contrastive,
  title={A contrastive learning approach for training variational autoencoder priors},
  author={Aneja, Jyoti and Schwing, Alex and Kautz, Jan and Vahdat, Arash},
  journal={Advances in neural information processing systems},
  volume={34},
  pages={480--493},
  year={2021}
}

@article{lucas2019understanding,
  title={Understanding posterior collapse in generative latent variable models},
  author={Lucas, James and Tucker, George and Grosse, Roger and Norouzi, Mohammad},
  year={2019}
}

@inproceedings{li2020system,
  title={{A System for Massively Parallel Hyperparameter Tuning}},
  author={Li, Liam and others},
  booktitle={Proc. of MLSys}, 
  year={2020}
}

@inproceedings{chang2022maskgit,
  title={{MaskGIT: Masked Generative Image Transformer}},
  author={Chang, Huiwen and Zhang, Han and Jiang, Lu and Liu, Ce and Freeman, William T},
  booktitle={Proc. of CVPR}, 
  year={2022}
}

@article{ctabgan+,
  title={{CTAB-GAN+: Enhancing Tabular Data Synthesis}},
  author={Zhao, Zilong and Kunar, Aditya and Birke, Robert and Van der Scheer, Hiek and Chen, Lydia Y},
  journal={Frontiers in big Data}, 
  year={2024} 
}

@inproceedings{gulrajani2017improved,
  title={{Improved Training of Wasserstein GANs}},
  author={Gulrajani, Ishaan and Ahmed, Faruk and Arjovsky, Martin and Dumoulin, Vincent and Courville, Aaron C},
  booktitle={Proc. of NeurIPS}, 
  year={2017}
}

@inproceedings{lee2023codi,
  title={{CoDi: Co-evolving Contrastive Diffusion Models for Mixed-type Tabular Synthesis}},
  author={Lee, Chaejeong and Kim, Jayoung and Park, Noseong},
  booktitle={Proc. of ICML}, 
  year={2023} 
}

@article{watanabe2023tree,
  title={{Tree-Structured Parzen Estimator: Understanding Its Algorithm Components and Their Roles for Better Empirical Performance}},
  author={Watanabe, Shuhei},
  journal={CoRR abs/2304.11127},
  year={2023}
}

@inproceedings{seedatcurated,
  title={{Curated LLM: Synergy of LLMs and Data Curation for tabular augmentation in low-data regimes}},
  author={Seedat, Nabeel and Huynh, Nicolas and van Breugel, Boris and van der Schaar, Mihaela},
  booktitle={Proc. of ICML},
  year={2023}
}

@inproceedings{kim2022stasy,
  title={{STaSy: Score-based Tabular Data Synthesis}},
  author={Kim, Jayoung and Lee, Chaejeong and Park, Noseong},
  booktitle={Proc. of ICLR},
  year={2023}
}

@inproceedings{kenton2019bert,
  title={BERT: Pre-training of Deep Bidirectional Transformers for Language Understanding},
  author={Kenton, Jacob Devlin Ming-Wei Chang and Toutanova, Lee Kristina},
  booktitle={Proc. of NAACL-HLT},
  year={2019}
}

@inproceedings{gulati2024tabmt,
  title={{TabMT: Generating Tabular data with Masked Transformers}},
  author={Gulati, Manbir and Roysdon, Paul},
  booktitle={Proc. of NeurIPS}, 
  year={2024}
}

@inproceedings{ctabgan,
  title={{CTAB-GAN: Effective Table Data Synthesizing}},
  author={Zhao, Zilong and Kunar, Aditya and Birke, Robert and Chen, Lydia Y},
  booktitle={Proc. of ACML}, 
  year={2021} 
}

@inproceedings{tabsyn,
  title={{Mixed-Type Tabular Data Synthesis with Score-Based Diffusion in Latent Space}},
  author={Zhang, Hengrui and others},
  booktitle={Proc. of ICLR},
  year={2024}
}

@article{solatorio2023realtabformer,
  title={REaLTabFormer: Generating Realistic Relational and Tabular Data using Transformers},
  author={Solatorio, Aivin V and Dupriez, Olivier},
  journal={Proc. of CoRR abs/2302.02041},
  year={2023}
}

@inproceedings{tvae/ctgan,
  title={Modeling tabular data using conditional gan},
  author={Xu, Lei and Skoularidou, Maria and Cuesta-Infante, Alfredo and Veeramachaneni, Kalyan},
  booktitle={Proc. of NIPS},
  year={2019}
}

@inproceedings{pierazzi2020intriguing,
  title={{Intriguing Properties of Adversarial ML Attacks in the Problem Space}},
  author={Pierazzi, Fabio and Pendlebury, Feargus and Cortellazzi, Jacopo and Cavallaro, Lorenzo},
  booktitle={Proc. of IEEE S\&P}, 
  year={2020} 
}

@inproceedings{stadler2022synthetic,
  title={{Synthetic Data – Anonymisation Groundhog Day}},
  author={Stadler, Theresa and Oprisanu, Bristena and Troncoso, Carmela},
  booktitle={Proc. of USENIX Security}, 
  year={2022}
}

@inproceedings{meghdouri2021controllable,
  title={{Controllable Network Data Balancing with GANs}},
  author={Meghdouri, Fares and Schmied, Thomas and G{\"a}rtner, Thomas and Zseby, Tanja},
  booktitle={Proc. of NeurIPS Workshop},
  year={2021}
}

@article{imbalanced-learn,
  title={{Imbalanced-learn: A Python Toolbox to Tackle the Curse of Imbalanced Datasets in Machine Learning}},
  author={{Lema{\~A}{\v{Z}}tre, Guillaume and Nogueira, Fernando and Aridas, Christos K}},
  journal={JMLR}, 
  year={2017}
}

@inproceedings{hao2021producing,
  title={{Producing More with Less: A GAN-based Network Attack Detection Approach for Imbalanced Data}},
  author={Hao, Xingran and Jiang, Zhengwei and Xiao, Qingsai and Wang, Qiuyun and Yao, Yepeng and Liu, Baoxu and Liu, Jian},
  booktitle={Proc. of CSCWD}, 
  year={2021} 
}

@inproceedings{li-etal-2024-search,
    title = "In Search of the Long-Tail: Systematic Generation of Long-Tail Inferential Knowledge via Logical Rule Guided Search",
    author = "Li, Huihan  and
     others",
    booktitle = "Proc. of EMNLP",
    year = "2024",
}

@inproceedings{khorram2024taming,
  title={Taming the Tail in Class-Conditional GANs: Knowledge Sharing via Unconditional Training at Lower Resolutions},
  author={Khorram, Saeed and Jiang, Mingqi and Shahbazi, Mohamad and Danesh, Mohamad H and Fuxin, Li},
  booktitle={Proc. of CVPR},
  year={2024}
}

@inproceedings{shao2024diffult,
title={Diffu{LT}: Diffusion for Long-tail Recognition Without External Knowledge},
author={Jie Shao and Ke Zhu and Hanxiao Zhang and Jianxin Wu},
booktitle={Proc. of NIPS},
year={2024}
}

@inproceedings{wang2024llmautoda,
title={{LLM}-Auto{DA}: Large Language Model-Driven Automatic Data Augmentation for Long-tailed Problems},
author={Pengkun Wang and Zhe Zhao and HaiBin Wen and Fanfu Wang and Binwu Wang and Qingfu Zhang and Yang Wang},
booktitle={Proc. of NIPS},
year={2024}
}

@inproceedings{park2025raretofrequent,
title={Rare-to-Frequent: Unlocking Compositional Generation Power of Diffusion Models on Rare Concepts with {LLM} Guidance},
author={Dongmin Park and others},
booktitle={Proc. of ICLR},
year={2025}
}

@inproceedings{tabddpm,
  title={{TabDDPM: Modelling Tabular Data with Diffusion Models}},
  author={Kotelnikov, Akim and Baranchuk, Dmitry and Rubachev, Ivan and Babenko, Artem},
  booktitle={Proc. of ICML}, 
  year={2023} 
}

@inproceedings{GreaT,
  title={{Language Models are Realistic Tabular Data Generators}},
  author={Borisov, Vadim and Se{\ss}ler, Kathrin and Leemann, Tobias and Pawelczyk, Martin and Kasneci, Gjergji},
  booktitle={Proc. of ICLR},
  year={2023}
}

@misc{pytorch-vision-transforms,
  title        = {PyTorch Vision Transforms},
  author       = {{PyTorch Contributors}},
  year         = {2023},
  howpublished = {\url{https://pytorch.org/vision/0.15/transforms.html}},
  note         = {Accessed: 2025-02-14}
}

@inproceedings{sabnis2021tragen,
  title={{TRAGEN: A Synthetic Trace Generator for Realistic Cache Simulations}},
  author={Sabnis, Anirudh and Sitaraman, Ramesh K},
  booktitle={Proc. ACM IMC}, 
  year={2021}
}

@inproceedings{bahnasy2020deepbgp,
  title={{DeepBGP: A Machine Learning Approach for BGP Configuration Synthesis}},
  author={Bahnasy, Mahmoud and Li, Fenglin and Xiao, Shihan and Cheng, Xiangle},
  booktitle={Proc. of NetAI Workshop}, 
  year={2020}
}

@article{xu2020toward,
  title={{Toward Effective Intrusion Detection Using Log-Cosh Conditional Variational Autoencoder}},
  author={Xu, Xing and Li, Jie and Yang, Yang and Shen, Fumin},
  journal={IEEE IoT}, 
  year={2020} 
}

@inproceedings{xu2021stan,
  title={{STAN: Synthetic Network Traffic Generation with Generative Neural Models}},
  author={Xu, Shengzhe and others},
  booktitle={Deployable Machine Learning for Security Defense: Second International Workshop, MLHat 2021, Virtual Event, August 15, 2021, Proceedings 2}, 
  year={2021} 
}

@incollection{aggarwal2014active, 
title={{Active Learning: A Survey}}, 
author={Aggarwal, Charu C and Kong, Xiangnan and Gu, Quanquan and Han, Jiawei and Philip, S Yu}, 
booktitle={Data Classification}, 
year={2014}, 
publisher={Chapman and Hall/CRC}
}

@article{ma2022comprehensive,
  title={{A Comprehensive Survey of Data Augmentation in Visual Reinforcement Learning}},
  author={Ma, Guozheng and Wang, Zhen and Yuan, Zhecheng and Wang, Xueqian and Yuan, Bo and Tao, Dacheng},
  journal={CoRR abs/2210.04561},
  year={2022}
}

@article{geurts2006extremely,
  title={{Extremely Randomized Trees}},
  author={Geurts, Pierre and Ernst, Damien and Wehenkel, Louis},
  journal={Machine learning}, 
  year={2006} 
}

\appendix
\section{Ethical Considerations}
\label{sec:ethics}
\noindent We used data and models that have been publicly shared for research purposes. We did not use human subjects in our research. Team members were not exposed to sensitive, private information, or disturbing content. All experiments were conducted in controlled lab settings and no deployed services were affected by our newly trained classifiers. For the study investigating data challenges, we collected titles and abstracts of papers published between 2017 and 2021 from publicly available proceedings on the USENIX and ACM websites, ensuring full compliance with their respective terms and conditions. 
The proposed/studied GenAI schemes generate synthetic data in the feature space (used by the defense schemes) and not in the problem space. We do not propose any methods for translating the data from the feature space to the problem space, and doing so would be challenging. Therefore, this limits any potential dual-use concerns of such GenAI technology by adversaries. 
However, an adversary can use GenAI schemes to generate training data to build a surrogate classifier, which can potentially be used to craft adversarial samples against the defense classifier. However, we believe that the benefits of the work towards cybersecurity defenses outweigh any potential harm caused by our study. 

\section{Paper selection methodology}
\label{appendix:measurement-paper-selection-method}
\noindent
In a semi-automated process, we shortlisted 35 cybersecurity papers (from top conferences) that use ML-based classifiers for network and application security tasks. We scraped titles and abstracts of 1,254 papers from USENIX Security and IEEE S\&P (2017–2021), then ranked them using latent semantic indexing~\cite{deerwester1990indexing} based on manually curated keywords (\eg \textit{classifier}, \textit{prediction}, \textit{accuracy}, \textit{epoch}). We manually reviewed top-ranked papers, selecting 25 that used security classifiers. An additional 10 papers came from team member suggestions.

\begin{table}[h!]
\centering
\setlength{\tabcolsep}{1pt}
\setlength\extrarowheight{3pt}
\begin{tabular}{l"l}
\multicolumn{1}{c"}{\textbf{\begin{tabular}[c]{@{}c@{}} Domain\end{tabular}}} & \multicolumn{1}{c}{\textbf{Papers with task descriptions}} \\ \Xhline{1.1pt}
\textbf{\begin{tabular}[c]{@{}l@{}} Network\\ security\end{tabular}} &
   \begin{tabular}[c]{p{0.82\columnwidth}}\raggedright Network intrusion detection system~\cite{sikder20176thsense,mirsky5kitsune}, infections with domain generation algorithm~\cite{schuppen2018fanci}, botnets~\cite{jan2020throwing, zhu2021botspot++}, IoT cryptojacking ~\cite{tekiner2022lightweight}, telephony scams~\cite{peeters2018sonar, li2018machine}, BGP serial hijackers~\cite{testart2019profiling}, malicious traffic~\cite{holland2021new, yu2020you}, abnormal cyber-physical patterns~\cite{chen2018learning}, and covert tunneling~\cite{barradas2018effective}.\end{tabular} \\ \hline
\textbf{\begin{tabular}[c]{@{}l@{}}Social  \\ media\\ \& web \\ security\end{tabular}} &
  \begin{tabular}[c]{p{0.82\columnwidth}}\raggedright Impersonation~\cite{cidon2019high}, phishing emails~\cite{ho2019detecting}, abusive social media accounts~\cite{xu2021deep}, stock pump-and-dump schemes~\cite{xu2019anatomy}, web traffic/ad fraud~\cite{yang2022wtagraph}, malicious URLs~\cite{thirumuruganathan2022siraj, chen2021cost}, website fingerprinting~\cite{gong2022surakav}, GDPR violations in cookies~\cite{bollinger2022automating}, and security vs. non-security browser crash causes~\cite{park2019empirical}.\end{tabular} \\ \hline
\textbf{\begin{tabular}[c]{@{}l@{}}Software \\ security\end{tabular}} &
   \begin{tabular}[c]{p{0.82\columnwidth}}\raggedright Malicious files~\cite{chen2020training}, executable binary code~\cite{jordaney2017transcend, downing2021deepreflect, yang2021cade, barbero2022transcending, yang2021bodmas}, software crash analysis~\cite{guo2019deepvsa}, resilience against automated attacks~\cite{banescu2017predicting, han2021sigl, she2020neutaint}, and vulnerability detection~\cite{she2019neuzz, zhou2019devign}.\end{tabular}
\end{tabular}
\caption{Measurement study papers with task categories.}
\label{data-challenges-tasks}
\end{table}

\begin{table}[h!]
\centering
\small
\setlength{\tabcolsep}{2pt}
\setlength\extrarowheight{2pt}
\begin{tabular}{c"cccccccc}
                                     & \multicolumn{8}{c}{\textbf{$\Delta G$ for months}}                                                                                                                                                                                                                                                                                                                                                  \\ \cline{2-9} 
\multirow{-2}{*}{\textbf{Technique}} & \multicolumn{2}{c|}{\textbf{1}}                                                                & \multicolumn{2}{c|}{\textbf{4}}                                                                & \multicolumn{2}{c|}{\textbf{8}}                                                                 & \multicolumn{2}{c}{\textbf{12}}                                                                \\ \Xhline{1.1pt}
\textbf{\nbase}                      & \multicolumn{1}{c|}{-8.3 (3.9)}   & \multicolumn{1}{c|}{\cellcolor[HTML]{DA5552}$\downarrow$} & \multicolumn{1}{c|}{-5.4 (7.1)}   & \multicolumn{1}{c|}{\cellcolor[HTML]{CCCCCC}-}            & \multicolumn{1}{c|}{-12.1 (10.1)}  & \multicolumn{1}{c|}{\cellcolor[HTML]{DA5552}$\downarrow$} & \multicolumn{1}{c|}{-7.6 (6.7)}    & {\cellcolor[HTML]{DA5552}$\downarrow$} \\ \hline
\textbf{\nrandom}                    & \multicolumn{1}{c|}{-8.8 (3.6)}   & \multicolumn{1}{c|}{\cellcolor[HTML]{DA5552}$\downarrow$} & \multicolumn{1}{c|}{6.5 (7.4)}    & \multicolumn{1}{c|}{\cellcolor[HTML]{CCCCCC}-}            & \multicolumn{1}{c|}{6.0 (9.2)}     & \multicolumn{1}{c|}{\cellcolor[HTML]{CCCCCC}-}            & \multicolumn{1}{c|}{-12.5 (9.1)}   & {\cellcolor[HTML]{DA5552}$\downarrow$} \\ \hline
\textbf{\tvae}                       & \multicolumn{1}{c|}{-12.5 (5.3)}  & \multicolumn{1}{c|}{\cellcolor[HTML]{DA5552}$\downarrow$} & \multicolumn{1}{c|}{-3.3 (10.1)}  & \multicolumn{1}{c|}{\cellcolor[HTML]{CCCCCC}-}            & \multicolumn{1}{c|}{-4.8 (18.6)}   & \multicolumn{1}{c|}{\cellcolor[HTML]{CCCCCC}-}            & \multicolumn{1}{c|}{-35.6 (10.2)}  & {\cellcolor[HTML]{DA5552}$\downarrow$} \\ \hline
\textbf{\tabsyn}                     & \multicolumn{1}{c|}{-11.4 (4.7)}  & \multicolumn{1}{c|}{\cellcolor[HTML]{DA5552}$\downarrow$} & \multicolumn{1}{c|}{-16.8 (9.5)}  & \multicolumn{1}{c|}{\cellcolor[HTML]{DA5552}$\downarrow$} & \multicolumn{1}{c|}{-25.2 (6.0)}   & \multicolumn{1}{c|}{\cellcolor[HTML]{DA5552}$\downarrow$} & \multicolumn{1}{c|}{-34.6 (14.6)}  & {\cellcolor[HTML]{DA5552}$\downarrow$} \\ \hline
\textbf{\ctabgan}                    & \multicolumn{1}{c|}{-4.9 (4.4)}   & \multicolumn{1}{c|}{\cellcolor[HTML]{DA5552}$\downarrow$} & \multicolumn{1}{c|}{3.5 (10.0)}   & \multicolumn{1}{c|}{\cellcolor[HTML]{CCCCCC}-}            & \multicolumn{1}{c|}{3.6 (14.9)}    & \multicolumn{1}{c|}{\cellcolor[HTML]{CCCCCC}-}            & \multicolumn{1}{c|}{-8.3 (15.2)}   & {\cellcolor[HTML]{CCCCCC}-}            \\ \hline
\textbf{\tabddpm}                    & \multicolumn{1}{c|}{-2.5 (0.8)}   & \multicolumn{1}{c|}{\cellcolor[HTML]{DA5552}$\downarrow$} & \multicolumn{1}{c|}{-2.5 (0.8)}   & \multicolumn{1}{c|}{\cellcolor[HTML]{DA5552}$\downarrow$} & \multicolumn{1}{c|}{-1.4 (4.4)}    & \multicolumn{1}{c|}{\cellcolor[HTML]{CCCCCC}-}            & \multicolumn{1}{c|}{-12.1 (7.1)}   & {\cellcolor[HTML]{DA5552}$\downarrow$} \\ \hline
\textbf{\smote}                      & \multicolumn{1}{c|}{-2.9 (1.1)}   & \multicolumn{1}{c|}{\cellcolor[HTML]{DA5552}$\downarrow$} & \multicolumn{1}{c|}{-2.9 (2.1)}   & \multicolumn{1}{c|}{\cellcolor[HTML]{DA5552}$\downarrow$} & \multicolumn{1}{c|}{-6.0 (5.2)}    & \multicolumn{1}{c|}{\cellcolor[HTML]{DA5552}$\downarrow$} & \multicolumn{1}{c|}{-9.6 (7.4)}    & {\cellcolor[HTML]{DA5552}$\downarrow$} \\ \hline
\textbf{\mcccr}                      & \multicolumn{1}{c|}{-1.5 (0.6)}   & \multicolumn{1}{c|}{\cellcolor[HTML]{DA5552}$\downarrow$} & \multicolumn{1}{c|}{-3.4 (1.6)}   & \multicolumn{1}{c|}{\cellcolor[HTML]{DA5552}$\downarrow$} & \multicolumn{1}{c|}{-5.7 (5.9)}    & \multicolumn{1}{c|}{\cellcolor[HTML]{CCCCCC}-}            & \multicolumn{1}{c|}{-19.5 (14.4)}  & {\cellcolor[HTML]{DA5552}$\downarrow$} \\ 
\end{tabular}
\caption{
Mean $\Delta G$ with standard deviation for \bodmas task for 1, 4, 8, \& 12 months. $\downarrow$ indicates negative gains. `-' indicates unreliable gains (CV $\geq$ 1). None of the methods show positive $\Delta G$ in these months. \great and \realtabformer's failures are not shown.}
\label{tab:bodmas-noeval-months}
\end{table}

\begin{table}[h!]
\centering
\small
\setlength{\tabcolsep}{2pt}
\setlength\extrarowheight{2pt}
\begin{tabular}{c"c|c|c|c|c|c|c}
\textbf{Task} & \textbf{TVAE} & \textbf{CT} & \textbf{TD} & \textbf{TabSyn} & \textbf{GReaT} & \textbf{RT} & \textbf{\sysname} \\ \Xhline{1.1pt}
\textbf{BGP} & 1 & 8.2 & 0.004 & 18.6 & 5.9 & 14 & 0.6 \\
\textbf{Tor} & 2 & 168 & 0.158 & 209 & 0.92 d* & 1041 & 9.46 \\
\textbf{IoT} & 16.6 & 837 & 0.27 & 652 & 238 d* & 2.7 h* & 72 \\
\textbf{Cookie} & 77.5 & 2,074 & 0.223 & 51 & $\times$ & 1.45 h* & 16.5 \\
\textbf{MS} & 10 & 206 & 0.01 & $\times$ & $\times$ & $\times$ & 3.57 \\
\textbf{BODMAS} & 3.9 & 232 & 0.123 & 14.7 & $\times$ & $\times$ & 1 \\
\textbf{nPrintML} & 59.4 & 2,368 & 0.047 & $\times$ & $\times$ & $\times$ & 21.8 \\ \Xhline{1.1pt}
\textbf{Avg. time} & 24.3 & 841.8 & 0.119 & - & - & - & 17.84
\end{tabular}%
\caption{Training times per epoch for GenAI schemes on an NVIDIA A100 GPU (in seconds unless marked as h = hours or d = days). $\times$ indicates model failure; * indicates estimated time. CT, TD, and RT refer to \ctabgan, \tabddpm, and \realtabformer, respectively.}
\label{tab:genai-train-times}
\end{table}

\section{Implementation of GenAI/non-GenAI tools}
\label{subsec:genai-additional}
\noindent This section discusses the hyper-parameters and implementation details of our chosen GenAI/non-GenAI tools.

\para{\tvae~\cite{tvae/ctgan}} We use Synthetic Data Vault (SDV)~\cite{sdv} to implement \tvae with early stopping. The model has 7 hyper-parameters: latent space size, encoder/decoder units, learning rate, batch size, loss factor, and number of epochs. As the authors provide no tuning guidelines, we use default settings.

\para{\ctabgan~\cite{ctabgan+}}We implement early stopping on \ctabgan's official source code. \ctabgan has 7 hyper-parameters, which include number of classifier layers, size of classifier layers, batch size, number of training epochs, number of channels, learning rate, number of encoding sides. Similar to \tvae, the authors provide no guidelines on how to tune the hyper-parameters, so we use the default settings.

\para{\tabddpm~\cite{tabddpm}}We implement \tabddpm using the official codebase. It has 10 hyper-parameters, including learning rate, iterations, diffusion timesteps, batch size, and MLP layers. The authors recommend and use the Tree-Structured Parzen Estimator~\cite{watanabe2023tree} (TPE) for hyper-parameter search.

\para{\tabsyn~\cite{tabsyn}}We implement \tabsyn using the official code, adding early stopping to the VAE training (already built-in for the Diffusion model). \tabsyn's architecture has 7 hyper-parameters: for the VAE—token dimension, number of layers, transformer layer dimension, $\lambda$, and min/max $\beta$; for the Diffusion model—hidden dimension. We skip hyper-parameter search, as the authors claim their parameter settings generalize across feature sizes.

\para{\great~\cite{GreaT}}We use \great's Python package with distilGPT as the backend LLM. We add early stopping and increase the context window from 512 to 1024 for tasks with $\geq$100 features. \great has no tunable hyper-parameters.

\para{\realtabformer~\cite{solatorio2023realtabformer}}We implemented \realtabformer using their official Python package.

\para{\smote~\cite{smote}} We use the imbalanced-learn~\cite{imbalanced-learn} implementation of \smote, with the number of neighbors in the $k$-nearest neighbors algorithm set to the default of 5.

\para{\mcccr~\cite{mc-ccr}}We implemented \mcccr using their source code, and best recommended generation settings: \textit{proportional} selection and \textit{translation} cleaning strategy.

\section{DeepSMOTE methodology}
\label{sub-sec:deepsmote}
\noindent 
DeepSMOTE~\cite{dablain2022deepsmote} uses an encoder-decoder architecture and \smote like interpolation in the latent space. During training, feature space samples are encoded into the latent space, where they are permuted and decoded to learn varied sample generation. During generation, feature space samples are encoded to the latent space, where \smote is applied, and decoded to produce feature space synthetic samples.

\begin{table}[h!]
\centering
\small
\setlength{\tabcolsep}{3pt}
\setlength\extrarowheight{3pt}
\begin{tabular}{c|c|c|c|c|c|c|c}
\textbf{Task} & \textbf{\bgp} & \textbf{Tor} & \textbf{\iot} & \textbf{\cookie} & \textbf{MS} & \textbf{\bodmas} & \textbf{\nprintml} \\ \hline
\textbf{\begin{tabular}[c]{@{}c@{}}\# of\\ samples\end{tabular}} & 146 & 5,749 & 41,634 & 12,882 & 619 & 11,592 & 686
\end{tabular}
\caption{Number of synthetic samples added during augmentation to balance out class distribution for all tasks. Here, MS refers to \microsoft.}
\label{tab:synth-samples-added}
\end{table}

\section{MC-CCR methodology}
\label{sub-sec:mcccr-details}
\noindent Here we describe \mcccr's methodology in detail. \mcccr~\cite{mc-ccr} follows a two-step process. In the first step, a cleaning procedure is performed \ie the majority samples in the proximity of the minority samples are removed. In the second step, additional minority samples are added in spherical regions around each minority sample. \mcccr uses a weighting strategy similar to ADASYN~\cite{adasyn} to compute the positions of synthetic samples to be added around difficult samples. The augmentation radius (sphere) is determined by the energy or budget allocated.

\section{\sysname Hyper-parameter search methodology}
\label{subsec:nimai-hyper-search}
\noindent 
\sysname's hyper-parameter search is conducted in two stages: first for the VQVAE model, then for the MTM model. To speed up the process, we use a stratified 10k-sample subset for tasks with over 10k training samples. Table~\ref{tab:hyper-parameters} lists the architectural hyper-parameters and their search spaces. For VQVAE, ASHA identifies parameters that minimize codebook (embedding + commitment) loss on a validation set, excluding those causing codebook collapse. For MTM, ASHA minimizes cross-entropy loss on the validation set.

\begin{table}[h!]
\centering
\setlength{\tabcolsep}{3pt}
\setlength\extrarowheight{3pt}
\begin{tabular}{c"l|l}
\textbf{\begin{tabular}[c]{@{}c@{}}Component\end{tabular}} &
  \multicolumn{1}{c|}{\textbf{Hyper-parameter}} & 
  \textbf{Search space} \\ \Xhline{1.1pt}
\multirow{3}{*}{\begin{tabular}[c]{@{}c@{}}Encoder, \\ Decoder, \\ MTM\end{tabular}} &
  \begin{tabular}[c]{@{}l@{}}Number of multi-heads\\ in attention layer.\end{tabular} &
  \{2, 4, 8, 16\} \\ \cline{2-3} 
 &
  \begin{tabular}[c]{@{}l@{}}Number of neurons in\\ feed-forward layer.\end{tabular} &
  \begin{tabular}[c]{@{}l@{}}\{16, 32, 64, \\ 128, 256, 512\}\end{tabular} \\ \cline{2-3} 
 & \begin{tabular}[c]{@{}l@{}}Number of transformer\\ layers.\end{tabular}             & \{1, 2, 4\} \\ \hline
\multirow{6}{*}{VQVAE} &
  \begin{tabular}[c]{@{}l@{}}Number of vectors in \\ codebook.\end{tabular} &
  \begin{tabular}[c]{@{}l@{}}\{32, 64, 96, 128,\\ 256, 512, 1024\}\end{tabular} \\ \cline{2-3} 
 &
  \begin{tabular}[c]{@{}l@{}}Length of vectors \\ in the latent space.\end{tabular} &
  \begin{tabular}[c]{@{}l@{}}2\textasciicircum{}n where n\\ between (5, 49)\end{tabular} \\ \cline{2-3} 
 &
  \begin{tabular}[c]{@{}l@{}}Size of embedding \\ vectors in codebook.\end{tabular} &
  \begin{tabular}[c]{@{}l@{}}\{2 , 4, 6, 12,\\ 16, 24, 32, 48\}\end{tabular} \\ \cline{2-3} 
 & \begin{tabular}[c]{@{}l@{}}Weighting factor of \\ codebook loss.\end{tabular}       & (1, 50)     \\ \cline{2-3} 
 & \begin{tabular}[c]{@{}l@{}}Weighting factor of\\ reconstruction loss.\end{tabular}  & (1, 50)     \\ \cline{2-3} 
 & \begin{tabular}[c]{@{}l@{}}Decay loss factor in\\ VQVAE EMA quantizer.\end{tabular} & (0, 1)     
\end{tabular}
\caption{Architectural hyper-parameters used in \nimai's VQVAE, encoder/decoder and MTM model components.}
\label{tab:hyper-parameters}
\end{table}

\begin{figure}[htbp]
\centering
\begin{subfigure}[htbp]{0.95\columnwidth}
\centering 
\includegraphics[scale=0.7]{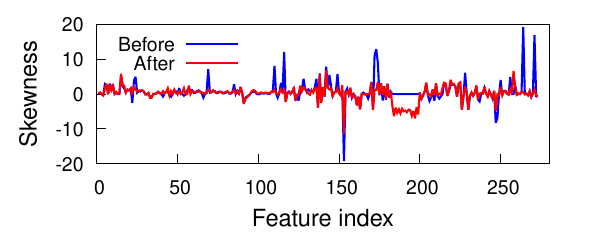}
\caption{\iot{}}
\end{subfigure}
\begin{subfigure}[t]{0.95\columnwidth}
\centering 
\includegraphics[scale=0.7]{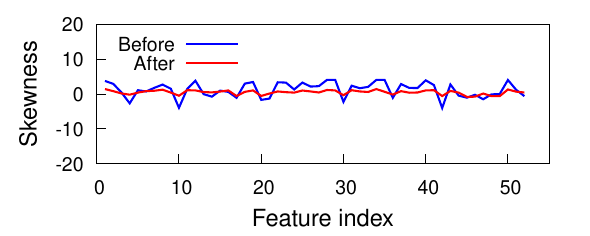}
\caption{\bgp{}}
\end{subfigure}
\caption{Feature-wise skewness comparison of attack-class samples before and after augmentation.}
\Description{A plot comparing the feature-wise skewness of attack-class samples before and after data augmentation. It shows changes in skewness values across multiple features, indicating how augmentation affects feature distributions.}
\label{fig:skewness-iot-bgp}
\end{figure}

\begin{table}[h!]
\centering
\small
\setlength{\tabcolsep}{3pt}
\setlength\extrarowheight{3pt}
\begin{tabular}{c"cc|c}
\multicolumn{1}{l"}{\multirow{2}{*}{\textbf{\begin{tabular}[c]{@{}l@{}} (N-1) \\ Month\end{tabular}}}} &
  \multicolumn{2}{c|}{\textbf{\begin{tabular}[c]{@{}c@{}}F-score for months \\ (w/ standard deviation)\end{tabular}}} &
  \multicolumn{1}{l}{\multirow{2}{*}{\textbf{\begin{tabular}[c]{@{}l@{}}Subset \\ size\end{tabular}}}} \\ \cline{2-3}
\multicolumn{1}{l"}{} & \multicolumn{1}{c|}{\textbf{5}}                       & \textbf{6}                        & \multicolumn{1}{l}{} \\ \Xhline{1.1pt}
\textbf{1}            & \multicolumn{1}{c|}{56.23 (4.55) / 35.4}          & 52.26 (2.77) / 33.8           & 34                   \\ \hline
\textbf{2}            & \multicolumn{1}{c|}{42.79 (2.81) / 36.4}          & 37.79 (4.01) / 33.4          & 49                   \\ \hline
\textbf{3}            & \multicolumn{1}{c|}{\textbf{59.78 (8.34) / 34.4}} & \textbf{53.92 (10.39) / 33.4} & 64                   \\ \hline
\textbf{4}            & \multicolumn{1}{c|}{50.11 (5.10) / 39}            & 44.13 (3.85) / 33.7          & 33                   \\ \hline
\textbf{5}            & \multicolumn{1}{c|}{}                                 & 48.27 (5.20) / 39.6          & 9                   
\end{tabular}
\caption{Mean macro F-scores w/ standard deviation for \sysname-hybrid separated (/) by macro F-scores for INSOMNIA on \bodmas months 5 \& 6. Best scores are bolded.}
\label{tab:concept-drift-all-months}
\end{table}

\section{Monthwise statistical tests on \bodmas}
\label{subsec:bodmas-stat-tests}
\noindent This section provides a month-by-month analysis of the 8 testing months where positive  gains are observed, presented in Table~\ref{tab:bodmas-months2356791011}. In \textbf{months 2, 3 and 11}, \nrandom is the only scheme that shows a positive gain over all other schemes (both GenAI and non-GenAI), further validating Finding 3 (Section~\ref{subsubsec:genai-potential}). In months 5, 6, 7, 9, and 10, several schemes show positive gains with high std values. Therefore, we apply statistical tests to compare the rankings of reliable schemes for these months.
 
In \textbf{months 5 and 6}, all GenAI schemes show positive gains. These months also experience the highest performance degradation under concept drift (Figure~\ref{fig:bodmas-test}). The Kruskal-Wallis test shows a significant difference in rankings in both months 5 and 6, with p-values of 1.044e-5 and 2.42e-3 ($<$ 0.05), respectively. 
Dunn's test for month 5 ranks the schemes as follows: \nrandom $>$ \ctabgan $>$ \tvae $>$ \nbase $>$ \tabddpm $>$ \tabsyn; clearly showing \nrandom to be the best performing scheme amongst all GenAI schemes with a percentage gain of 59.42\%. 
Dunn's test for month 6 ranks the schemes as follows: \tvae $>$ \ctabgan $>$ \nbase $>$ \nrandom $>$ \tabddpm $>$ \tabsyn, with \tvae outperforming all other schemes. Note, \nbase is the only sample-conditioned scheme to outperform \tabddpm, \tabsyn and \nrandom class-conditioned GenAI schemes. Traditional sample-conditioned schemes exhibit unreliable performance in both months. 

In \textbf{month 7}, only \nrandom, \tabddpm and \smote show positive gains. Kruskal-Wallis test shows a significant difference in means with a p-value of 0.0982 $<$ 0.05. Post-hoc tests rank the schemes as \smote $>$ \nrandom $>$ \tabddpm. Here, traditional sample- conditioned \smote method outperforms all class-conditioned schemes including \nrandom and \tabddpm. 

In \textbf{month 9}, \nrandom and \tvae show positive gains. In \textbf{month 10}, \nrandom, \nbase, and \tabddpm show positive gains. The results of Kruskal-Wallis test indicate that there is no difference in means for the schemes with positive gains in both months. p-values for months 9 and 10 are 0.409 and 0.239 ($>$0.05) respectively. Therefore, in month 9, \nrandom and \tvae are the top performers, and in month 10, \nrandom, \nbase and \tabddpm lead, all 3 of them being class-conditioned schemes.

\section{Impact of noisy class labels}
\label{subsec:noisy-bgp-effect}
\noindent We study how noisy/overlapping class samples affect data augmentation. In the original \bgp task, sample-conditioned schemes show positive gains (Table~\ref{tab:genai-success-results}). We induce class overlap by flipping 20\% of class-0 labels. Results for top-performing GenAI and traditional methods are in Table~\ref{tab:bgp-overlap-results}. All methods degrade under overlap, highlighting the limited robustness of current augmentation schemes, especially GenAI, to noisy/overlapping data.

\begin{table}[ht]
\centering
\small
\setlength{\tabcolsep}{4pt}
\setlength\extrarowheight{2pt}
\begin{tabular}{c"cc}
\multirow{2}{*}{\textbf{Technique}} & \multicolumn{2}{c}{\textbf{\begin{tabular}[c]{@{}c@{}}Mean $\Delta G$ \end{tabular}}} \\ \cline{2-3} 
 & \multicolumn{2}{c}{\textbf{\bgp Overlap}} \\ \Xhline{1.1pt}
\textbf{Real} & \multicolumn{1}{c|}{40.50 (-)} &   \\ \hline
\textbf{\nbase} & \multicolumn{1}{c|}{-14.02 (2.26)} & \cellcolor[HTML]{da5552}$\downarrow$ \\ \hline
\textbf{\nrandom} & \multicolumn{1}{c|}{-0.46 (2.28)} & \cellcolor[HTML]{da5552}$\downarrow$ \\ \hline
\textbf{\ctabgan} & \multicolumn{1}{c|}{-25.35 (2.54)} & \cellcolor[HTML]{da5552}$\downarrow$ \\ \hline
\textbf{\smote} & \multicolumn{1}{c|}{-37.40 (3.99)} & \cellcolor[HTML]{da5552}$\downarrow$ \\ 
\end{tabular}
\caption{Mean $\Delta G$ with standard deviation for overlapping class data setting in \bgp. $\downarrow$ indicates negative gains.}
\label{tab:bgp-overlap-results}
\end{table}

\begin{table}[h!]
\centering
\small
\setlength{\tabcolsep}{3pt}
\setlength\extrarowheight{3pt}
\begin{tabular}{cccccccccccc}
\multicolumn{12}{c}{\textbf{\# Test samples for \bodmas months}} \\ \hline
\multicolumn{1}{c|}{\textbf{1}} &
  \multicolumn{1}{c|}{\textbf{2}} &
  \multicolumn{1}{c|}{\textbf{3}} &
  \multicolumn{1}{c|}{\textbf{4}} &
  \multicolumn{1}{c|}{\textbf{5}} &
  \multicolumn{1}{c|}{\textbf{6}} &
  \multicolumn{1}{c|}{\textbf{7}} &
  \multicolumn{1}{c|}{\textbf{8}} &
  \multicolumn{1}{c|}{\textbf{9}} &
  \multicolumn{1}{c|}{\textbf{10}} &
  \multicolumn{1}{c|}{\textbf{11}} &
  \textbf{12} \\ \hline
\multicolumn{1}{c|}{826} &
  \multicolumn{1}{c|}{829} &
  \multicolumn{1}{c|}{1563} &
  \multicolumn{1}{c|}{1205} &
  \multicolumn{1}{c|}{623} &
  \multicolumn{1}{c|}{1087} &
  \multicolumn{1}{c|}{945} &
  \multicolumn{1}{c|}{1073} &
  \multicolumn{1}{c|}{283} &
  \multicolumn{1}{c|}{268} &
  \multicolumn{1}{c|}{312} &
  559
\end{tabular}
\caption{Test set statistics for \bodmas over 12 months. }
\label{tab:bodmas-test-samples.}
\end{table}

\section{\sysname{} Training Objective}
\label{appendix:training-objective-nimai}
\sysname{}'s training objectives include the following loss terms.

\textit{Reconstruction loss ($\mathcal{L}_{\text{recon}}$)}: This term ensures accurate reconstruction in feature space using mean squared error (MSE). For input $x$ and reconstruction $\hat{x}$:
\begin{equation}
    \mathcal{L}_{\text{recon}} = \| x - \hat{x} \|^2
    \label{equation:reconstruction}
\end{equation}

\textit{Commitment Loss ($\mathcal{L}_{\text{commit}}$)}: This aligns encoder outputs with the nearest codebook embeddings, where \( z_e(x) \) is the encoder output, \( e \) the closest embedding, \( sg(\cdot) \) the stop-gradient operator, and \( \beta \) a scaling term:
\begin{equation}
    \mathcal{L}_{\text{commit}} = \beta  \| z_e(x) - \text{sg}(e) \|^2
    \label{equation:commitment}
\end{equation}

\textit{Embedding Loss ($\mathcal{L}_{\text{embed}}$)}: This updates embedding vectors to match encoder outputs:
\begin{equation}
    \mathcal{L}_{\text{embed}} =  \| \text{sg}(z_e(x)) - e \|^2
    \label{equation:embedding}
\end{equation}

Finally, the overall objective combines all losses, weighted by \( \alpha \):
\begin{equation}
    \mathcal{L} = \mathcal{L}_{\text{recon}} + \alpha \left( \mathcal{L}_{\text{embed}} + \mathcal{L}_{\text{commit}} \right)
    \label{equation:vqvae-loss}
\end{equation}

\section{MTM Loss Function and Training Objective}
\label{appendix:mtm-objective}

MTM's objective in Equation~\ref{equation:maskgit} uses cross-entropy loss to estimate the log-likelihood of quantized latent tokens. Here, \( \mathbf{y}_i \) is a latent feature, \( \mathbf{Y}_{\text{mask}} \) a masked token filled by MTM, and \( \mathbf{c} \) the class label.

\begin{equation}
\mathcal{L}_{\text{MTM}} = \mathbb{E}_{(\mathbf{y}, c) \sim D} \left[ - \sum_{i} \log p\left( \mathbf{y}_i \mid \mathbf{Y}_{\text{mask}}, c \right) \right]
\label{equation:maskgit}
\end{equation}

\end{document}